\documentclass[10pt,twocolumn,aps,prd,preprintnumbers,showpacs,superscriptaddress,nofootinbib,amsmath,amssymb,floats,floatfix,showkeys,notitlepage,longbibliography,fleqn]{revtex4-2}
\usepackage{orcidlink}
\usepackage{comment}
\usepackage{lipsum}
\usepackage{graphicx}
\usepackage{subfigure}
\usepackage{palatino}
\usepackage{sans}
\usepackage{hyperref}
\hypersetup{colorlinks=true,linkcolor=blue,urlcolor=blue,citecolor=blue}
\usepackage[toc,page]{appendix}
\usepackage[normalem]{ulem}
\usepackage{adjustbox}
\usepackage{latexsym}
\usepackage{amsmath}
\usepackage{amssymb}
\usepackage{amsfonts}
\usepackage{dcolumn}
\usepackage{bm}
\usepackage{tikz}
\usetikzlibrary{decorations.pathmorphing}
\usepackage{bigints}
\usepackage{array,tabularx,multirow,booktabs}
\usepackage[tracking=true]{microtype}
\usepackage{soul} 
\SetTracking{}{500}
\SetTracking{encoding={*}, shape=sc}{40}
\UseRawInputEncoding 
\allowdisplaybreaks
\usepackage[utf8]{inputenc}
\usepackage{xcolor} 
\newcommand{\nn}{\nonumber \\}

\begin{document} \sloppy

\title{Reference-renormalized Jacobi primitives for finite-distance massive-particle deflection}

\author{Reggie C. Pantig \orcidlink{0000-0002-3101-8591}} 
\email{rcpantig@mapua.edu.ph}
\affiliation{Physics Department, School of Foundational Studies and Education, Map\'ua University, 658 Muralla St., Intramuros, Manila 1002, Philippines.}

\begin{abstract}
We develop a reference-renormalized Jacobi-primitive Gauss-Bonnet formalism for finite-distance weak deflection of massive particles in static, spherically symmetric spacetimes.  The construction uses the fixed-energy Jacobi metric associated with timelike geodesics and treats the radial curvature primitive as a quantity defined only up to an additive constant.  Instead of fixing this constant by requiring the primitive to vanish at a circular timelike orbit, we subtract a reference Jacobi geometry at the same conserved particle energy and impose endpoint normalization on the curvature-density discrepancy.  This gives a circular-orbit-free primitive representation of the Gauss-Bonnet curvature contribution while preserving the operational finite-distance definition of the bending angle through local source and receiver directions.  We prove that the reference-normalized and circular-orbit-normalized prescriptions agree whenever both are admissible, since their primitives differ only by a constant that cancels from the exact curvature strip integral.  The method remains applicable when no admissible circular timelike orbit exists, when the orbit lies outside the Riemannian Jacobi domain, or when the natural no-lens comparison geometry is non-Euclidean.  We demonstrate the formalism in three representative settings.  For the Schwarzschild-like bumblebee spacetime, the method separates the conical reference contribution from the localized mass correction and recovers the Schwarzschild massive-particle limit when the deformation is removed.  For the Kottler spacetime, a de Sitter Jacobi reference organizes the localized mass contribution inside the static patch. For the Janis-Newman-Winicour geometry, the reference-renormalized primitive remains well defined in a photon-sphere-free JNW subinterval $0<\gamma\le2/5$ where unbound circular-orbit normalization is unavailable, and the finite-distance bending angle is completed through second order with explicit dependence on the scalar parameter at that order. The formalism therefore identifies circular-orbit anchoring as an auxiliary primitive gauge choice rather than a prerequisite for finite-distance Jacobi Gauss-Bonnet deflection.
\end{abstract}


\maketitle

\section{Introduction}\label{sec_introduction}
Gravitational deflection is most often discussed for light rays, but massive particles also define a sharp scattering problem whose bending depends on conserved energy, endpoint position, and local observer frame.  In finite-distance lensing, the source and receiver are not sent to an idealized infinity, so the observable must be defined by local directions measured at the endpoints and by the coordinate angular separation of the trajectory.  This viewpoint is essential not only for realistic source-receiver configurations, but also for backgrounds in which an asymptotically Euclidean region is absent or not the natural comparison geometry.  The finite-distance Gauss-Bonnet approach provides a geometric formulation of this problem by expressing the bending angle in terms of curvature and boundary contributions on an auxiliary two-dimensional geometry \cite{Gibbons:2008rj,Ishihara:2016vdc,Ono:2017pie,Takizawa:2020egm,Huang:2023bto}.

For null rays in a static spacetime, the relevant auxiliary geometry is the optical metric.  For massive particles, the corresponding object is the fixed-energy Jacobi metric, whose geodesics reproduce the spatial projections of timelike geodesics \cite{Gibbons:2015qja,Arganaraz:2021fwu}.  This distinction is important: the Jacobi metric depends explicitly on the conserved particle energy, and its Riemannian domain is restricted by the condition that the fixed-energy conformal factor remain positive.  The Jacobi formulation nevertheless gives a natural route to a Gauss-Bonnet description of massive-particle deflection, and it has been used to compute weak bending for static and stationary geometries, including black holes, wormholes, Lorentz-violating backgrounds, loop-quantum-gravity-inspired black holes, and generalized finite-domain GW constructions \cite{Crisnejo:2018uyn,Chanda:2019guf,Li:2019qyb,Li:2020dln,Huang:2022iwl,Huang:2023bto}.

A particularly efficient implementation was introduced by Li, Zhang, and \"Ovg\"un, who used circular orbits of the optical or Jacobi metric to reduce the curvature-area contribution to a one-dimensional primitive evaluated along the particle trajectory \cite{Li:2020wvn}.  Related single-integral and finite-domain versions of the Gibbons-Werner method have also emphasized that the practical reduction of the curvature contribution depends on how the integration domain and its boundary are chosen \cite{Huang:2022iwl,Huang:2023bto}.  In the circular-orbit method, the radial primitive of the curvature density is fixed by requiring it to vanish at an appropriate circular orbit.  This is geometrically economical when such an orbit exists in the relevant domain, but the normalization is not universal.  A circular timelike orbit may be absent at the scattering energy, may lie outside the Riemannian Jacobi region, may be hidden behind a horizon or singularity, or may be disconnected from the reference geometry needed to formulate a finite-distance observable in a non-Euclidean background.

The central point of the present work is that the circular orbit is not part of the definition of the finite-distance bending angle.  It fixes only the additive representative of a curvature primitive.  The primitive itself is defined modulo a constant, whereas the exact Gauss-Bonnet curvature strip depends only on primitive differences.  We therefore replace the circular-orbit normalization by a reference normalization: we subtract a reference Jacobi geometry constructed at the same conserved particle energy and impose that the resulting discrepancy primitive vanish at a chosen matching end.  This gives a circular-orbit-free Jacobi-primitive formalism while preserving the operational finite-distance definition through local source and receiver angles. A reference-renormalized curvature-primitive formulation has recently been developed for finite-distance weak deflection of light in static spherical spacetimes \cite{Pantig:2026xjj}.  The present work transfers the same normalization principle to massive-particle scattering, where the energy dependence and Riemannian-domain restriction of the Jacobi metric introduce additional structure that is absent in the null optical case.

The reference geometry plays two roles that must be kept distinct.  First, it fixes the additive ambiguity of the radial curvature primitive.  Second, when the background is not asymptotically Euclidean, it specifies the no-lens comparison geometry relative to which the finite-distance bending is organized.  In asymptotically flat scattering, the natural reference is the Minkowski Jacobi geometry at the same asymptotic speed.  In Kottler-type spacetimes, the natural reference is instead the de Sitter Jacobi geometry at the same conserved energy inside the static patch, in parallel with the broader literature on massive-particle deflection in Schwarzschild-de Sitter geometry \cite{He:2020eah}.  In conical geometries, such as the Schwarzschild-like bumblebee spacetime, the reference is flat but not Euclidean in the angular sector, so the closure contribution must be retained rather than discarded \cite{Li:2020dln,Li:2020wvn}.

We prove that the reference-normalized and circular-orbit-normalized prescriptions agree whenever both are admissible.  The reason is simple but important: two primitives of the same Jacobi curvature density differ only by a constant on a connected regular domain, and this constant cancels from the exact curvature strip integral.  Thus the circular-orbit method is recovered in its domain of validity, while the reference-renormalized method remains meaningful outside that domain.  The new prescription is therefore not a different observable; it is a more flexible normalization of the same Gauss-Bonnet curvature contribution.

The applications are chosen to demonstrate this point without spending equation budget on a standalone Schwarzschild example.  We first treat the Schwarzschild-like bumblebee spacetime, where the reference geometry is conical and the result can be compared directly with the circular-orbit Jacobi prescription.  We then treat the Kottler spacetime using a de Sitter Jacobi reference, which displays how finite-distance mass-cosmological terms are organized when the no-lens background is not Minkowski.  Finally, we consider the Janis-Newman-Winicour geometry as a naked-singularity example. In the photon-sphere-free JNW subinterval $0<\gamma\le2/5$ considered here, the circular timelike Jacobi orbit needed for unbound scattering normalization is unavailable, while the reference-renormalized primitive remains well defined and yields the finite-distance bending angle through second order.  The usual Schwarzschild massive-particle result appears only as a limiting check of these applications.

We restrict the analysis to static, spherically symmetric spacetimes and to equatorial, unbound, single-pass timelike trajectories.  This ensures that the Jacobi metric is Riemannian on the Gauss-Bonnet domain and that the projected orbit defines a simple boundary suitable for the primitive reduction.  We do not treat stationary axisymmetric spacetimes in this paper, since the massive-particle geometry is then generally Randers-Finsler rather than Riemannian and requires additional control of osculating metrics and geodesic-curvature terms \cite{Werner:2012rc,Chanda:2019guf,Li:2019qyb,Huang:2023bto}.  We also leave dispersive media, charged-particle external forces, strong deflection, and multiple-winding or bound-orbit domains for separate work \cite{Crisnejo:2019ril,Huang:2022gon}.

The paper is organized as follows.  Section \ref{sec_timelike_jacobi_observable} fixes the timelike scattering conventions, introduces the fixed-energy Jacobi metric, and defines the finite-distance observable.  Section \ref{sec_jacobi_reference_renormalization} derives the Jacobi Gauss-Bonnet primitive, introduces the reference-renormalized curvature discrepancy, and states the additive-gauge proposition that establishes equivalence with circular-orbit normalization.  Section \ref{sec_weak_expansion_protocol} gives the weak-deflection expansion protocol used in the applications.  Sections \ref{sec_bumblebee_application}, \ref{sec_kottler_application}, and \ref{sec_jnw_application} apply the method to the Schwarzschild-like bumblebee, Kottler, and Janis-Newman-Winicour geometries, respectively.  Section \ref{sec_discussion} discusses reference dependence, limiting cases, and possible extensions, and Sec. \ref{sec_conclusions} summarizes the results.  We use units $G=c=1$, metric signature $(-,+,+,+)$, and restrict all displayed Jacobi quantities to the equatorial plane.

\section{Timelike Jacobi scattering and the finite-distance observable}
\label{sec_timelike_jacobi_observable}
We fix the kinematical and geometrical conventions for massive-particle scattering in a static, spherically symmetric spacetime.  The purpose of this section is not to derive the field equations of any particular solution, but to isolate the fixed-energy Jacobi geometry on which the Gauss-Bonnet construction will be applied.  The finite-distance observable is defined operationally by local endpoint directions and by the angular span of the projected timelike trajectory.

\subsection{Static spherical timelike motion}
\label{subsec_static_spherical_timelike_motion}
We consider a four-dimensional static and spherically symmetric metric written as
\begin{align}
ds^2=-A(r)dt^2+B(r)dr^2+C(r)d\Omega^2 ,
\nn
d\Omega^2=d\theta^2+\sin^2\theta\,d\phi^2 .
\label{eq_sss_metric_sec2}
\end{align}
The functions $A(r)$, $B(r)$, and $C(r)$ are assumed positive on the scattering region.  The condition $A(r)>0$ keeps the analysis inside a static domain, while $C(r)>0$ labels the symmetry two-spheres.  By spherical symmetry, the orbit may be restricted to the equatorial plane $\theta=\pi/2$.

Let $\tau$ denote proper time, and let an overdot denote differentiation with respect to $\tau$.  Stationarity and axial symmetry give the conserved energy and angular momentum per unit rest mass,
\begin{align}
e=A(r)\dot t,
\nn
\ell=C(r)\dot\phi .
\label{eq_conserved_e_l_sec2}
\end{align}
The normalization $u^\mu u_\mu=-1$ then gives the radial equation
\begin{align}
B(r)\dot r^2
=
\frac{e^2}{A(r)}-1-\frac{\ell^2}{C(r)}
=
\frac{H(r,e,\ell)}{A(r)C(r)},
\nn
H(r,e,\ell)\equiv C(r)\left[e^2-A(r)\right]-A(r)\ell^2 .
\label{eq_radial_equation_H_sec2}
\end{align}
The function $H(r,e,\ell)$ will be used repeatedly below.  A single-pass unbound orbit has one radial turning point $r=r_0$, determined by $H(r_0,e,\ell)=0$.  Equivalently,
\begin{equation}
\ell^2=\frac{C(r_0)\left[e^2-A(r_0)\right]}{A(r_0)} .
\label{eq_turning_point_sec2}
\end{equation}
Combining Eq. \eqref{eq_conserved_e_l_sec2} with Eq. \eqref{eq_radial_equation_H_sec2}, the projected orbit satisfies
\begin{equation}
\frac{d\phi}{dr}
=
\pm
\frac{\ell\sqrt{A(r)B(r)}}{\sqrt{C(r)}\sqrt{H(r,e,\ell)}} .
\label{eq_orbit_equation_sec2}
\end{equation}
The sign distinguishes the ingoing and outgoing radial branches.  We choose the azimuthal orientation so that $\phi$ increases from the source to the receiver.

When the spacetime is asymptotically flat, with $A(r)\to1$, $B(r)\to1$, and $C(r)\sim r^2$, it is useful to replace $(e,\ell)$ by the asymptotic speed $v$ and impact parameter $b$,
\begin{align}
e&=\frac{1}{\sqrt{1-v^2}},
\nn
\ell&=\frac{bv}{\sqrt{1-v^2}},
\nn
b&=\frac{\ell}{\sqrt{e^2-1}},
\label{eq_v_b_def_sec2}
\end{align}
with $0<v<1$. For non-asymptotically flat geometries we keep $(e,\ell)$ as the primary constants of motion.  The speed $v$ introduced in Eq. \eqref{eq_v_b_def_sec2} has a physical interpretation only when an asymptotically flat normalization is available.  We therefore do not interpret $v$ as a particle speed in a de Sitter static patch.  The conserved energy is $e=-u_\mu\xi^\mu$ with the static Killing generator $\xi=\partial_t$, so its numerical value depends on the normalization of $\xi$.  A fixed-energy reference subtraction consequently requires the physical and reference Killing generators to be identified before their Jacobi geometries are compared.  When the reference spacetime is obtained as a member of the same static family, we retain the same time coordinate and do not rescale the two Killing generators independently.

\subsection{Jacobi metric and local direction angle}
\label{subsec_jacobi_metric_local_angle}
At fixed conserved energy, the spatial projection of a timelike geodesic in a static spacetime is a geodesic of the Jacobi metric.  Up to an irrelevant overall factor equal to the squared rest mass, the equatorial Jacobi line element is 
\begin{equation}
d\ell_J^2
=
\left[e^2-A(r)\right]
\left[
\frac{B(r)}{A(r)}dr^2
+
\frac{C(r)}{A(r)}d\phi^2
\right] .
\label{eq_jacobi_metric_sec2}
\end{equation}
This is the standard fixed-energy Jacobi metric for timelike geodesics in static spacetimes, specialized here to the equatorial plane and to unit rest mass \cite{Gibbons:2015qja,Arganaraz:2021fwu}. The metric is Riemannian only in the fixed-energy region
\begin{align}
A(r)&>0,
\nn
B(r)&>0,
\nn
C(r)&>0,
\nn
e^2&>A(r).
\label{eq_jacobi_domain_sec2}
\end{align}
All Gauss-Bonnet domains used in this paper are assumed to lie inside this region.  The corresponding area density is
\begin{equation}
\sqrt{\det g_J}
=
\frac{e^2-A(r)}{A(r)}\sqrt{B(r)C(r)} .
\label{eq_jacobi_area_sec2}
\end{equation}

Along the physical timelike geodesic, the relation between proper time and Jacobi arclength follows directly from the normalization equation,
\begin{align}
\frac{d\ell_J}{d\tau}
&=
\frac{e^2-A(r)}{A(r)},
\nn
\frac{d\phi}{d\ell_J}
&=
\frac{A(r)\ell}{C(r)\left[e^2-A(r)\right]} .
\label{eq_jacobi_arclength_sec2}
\end{align}
Let $\Psi_J$ be the angle, measured with the Jacobi metric, between the particle direction and the outward radial direction.  In the orthonormal frame of $g_J$, Eq. \eqref{eq_jacobi_arclength_sec2} gives
\begin{equation}
\sin\Psi_J(r)
=
\ell
\sqrt{
\frac{A(r)}
{C(r)\left[e^2-A(r)\right]}
} .
\label{eq_local_angle_sec2}
\end{equation}
This is also the local direction angle measured by a static observer.  Indeed, the azimuthal component of the particle momentum in the static orthonormal frame is $\ell/\sqrt{C(r)}$, while its spatial momentum magnitude is $\sqrt{e^2/A(r)-1}$.  Their ratio is precisely the right-hand side of Eq. \eqref{eq_local_angle_sec2}.  Thus the Jacobi angle is not merely auxiliary; in the static spherical case it is the operational endpoint angle entering the finite-distance bending observable.

The null optical metric is recovered from the Jacobi metric by removing the divergent energy scale and taking the ultra-relativistic limit:
\begin{equation}
\lim_{e\to\infty}e^{-2}d\ell_J^2
=
\frac{B(r)}{A(r)}dr^2
+
\frac{C(r)}{A(r)}d\phi^2 .
\label{eq_null_limit_jacobi_sec2}
\end{equation}
This limit will be used as a consistency check for the reference-renormalized massive-particle formula.

\subsection{Finite-distance bending angle}
\label{subsec_finite_distance_observable_sec2}
Let the particle travel from a source point $S=(r_S,\phi_S)$ to a receiver point $R=(r_R,\phi_R)$ along an unbound trajectory with a single radial turning point.  We place the turning point at $\phi=0$, take the source on the ingoing branch, and take the receiver on the outgoing branch.  The coordinate angular span is $\phi_{RS}\equiv\phi_R-\phi_S$.

The finite-distance deflection angle is defined by
\begin{equation}
\alpha_J=\Psi_R-\Psi_S+\phi_{RS}.
\label{eq_alpha_def_sec2}
\end{equation}
This finite-distance form follows the local-angle definition used in the finite-distance Gauss-Bonnet literature, adapted here to the Jacobi metric for massive particles \cite{Ishihara:2016vdc,Ono:2017pie,Takizawa:2020egm,Li:2019qyb}. Here $\Psi_S$ and $\Psi_R$ are the local static-observer angles, equivalently the Jacobi angles of Eq. \eqref{eq_local_angle_sec2}, evaluated at the source and receiver.  For the branch convention just described,
\begin{align}
\Psi_R&=
\arcsin\!\left[
\ell
\sqrt{
\frac{A(r_R)}
{C(r_R)\left[e^2-A(r_R)\right]}
}
\right],
\nn
\Psi_S&=
\pi-
\arcsin\!\left[
\ell
\sqrt{
\frac{A(r_S)}
{C(r_S)\left[e^2-A(r_S)\right]}
}
\right].
\label{eq_psi_branches_sec2}
\end{align}
The endpoint angles are real only when
\begin{align}
\ell^2
\le
\frac{C(r_R)\left[e^2-A(r_R)\right]}{A(r_R)},
\nn
\ell^2
\le
\frac{C(r_S)\left[e^2-A(r_S)\right]}{A(r_S)} .
\label{eq_endpoint_admissibility_sec2}
\end{align}
These inequalities are the finite-distance kinematic admissibility conditions for a massive particle with constants $(e,\ell)$ to pass through the prescribed source and receiver radii.

The angular span is obtained by adding the two radial branches of Eq. \eqref{eq_orbit_equation_sec2},
\begin{align}
\phi_{RS}
&=
\int_{r_0}^{r_R}
\frac{\ell\sqrt{A(r)B(r)}\,dr}
{\sqrt{C(r)}\sqrt{H(r,e,\ell)}}
\nn &+
\int_{r_0}^{r_S}
\frac{\ell\sqrt{A(r)B(r)}\,dr}
{\sqrt{C(r)}\sqrt{H(r,e,\ell)}} .
\label{eq_phi_RS_integral_sec2}
\end{align}
Equations \eqref{eq_alpha_def_sec2}-\eqref{eq_phi_RS_integral_sec2} define the finite-distance observable that the Jacobi Gauss-Bonnet construction must reproduce.  The remaining formalism will not modify this definition.  It will only rewrite the curvature contribution to the same observable in a form whose primitive normalization does not require an auxiliary circular timelike orbit.

\section{Jacobi Gauss-Bonnet primitive and reference renormalization}
\label{sec_jacobi_reference_renormalization}
We now rewrite the finite-distance observable of Sec. \ref{sec_timelike_jacobi_observable} as a Gauss-Bonnet identity on the equatorial Jacobi manifold.  The essential simplification is that, for a static spherical metric, the Jacobi curvature density is a radial total derivative.  The corresponding primitive is defined only up to an additive constant.  We fix this constant by reference subtraction rather than by imposing a circular-orbit normalization.

\subsection{Gauss-Bonnet domain and curvature-density generator}
\label{subsec_gb_domain_curvature_generator}
Let $\mathcal{M}_J$ be the two-dimensional Jacobi manifold with coordinates $(r,\phi)$ and metric $g^J_{ab}dx^a dx^b=d\ell_J^2$.  The projected timelike trajectory is a Jacobi geodesic $\gamma_J$ joining the source $S$ to the receiver $R$.  We take a simply connected domain $D_J\subset\mathcal{M}_J$ bounded by $\gamma_J$, two radial segments through $S$ and $R$, and an outer closure curve $C_\Gamma$ in a reference or matching region.  The Gauss-Bonnet theorem gives \cite{Gibbons:2008rj,Ishihara:2016vdc,Li:2019qyb,Huang:2023bto}
\begin{equation}
\iint_{D_J}K_J\,dS_J
+
\oint_{\partial D_J}\kappa_J\,d\ell_J
+
\sum_i\beta_i
=
2\pi ,
\label{eq_GB_J_sec3}
\end{equation}
where $K_J$ is the Gaussian curvature of $g^J_{ab}$, $dS_J=\sqrt{\det g_J}\,dr\,d\phi$, $\kappa_J$ is the geodesic curvature of the boundary, and $\beta_i$ are exterior jump angles.

The physical path $\gamma_J$ contributes no geodesic curvature.  The radial boundaries are also unparameterized geodesics because the Jacobi metric is diagonal and depends only on $r$.  Hence
\begin{equation}
\Gamma^\phi_{rr}(g_J)=0,
\qquad
\oint_{\partial D_J}\kappa_J\,d\ell_J
=
\int_{C_\Gamma}\kappa_J\,d\ell_J .
\label{eq_radial_geodesic_boundary_sec3}
\end{equation}
With the orientation fixed by the source-to-receiver motion, the endpoint exterior angles are $\beta_S=\Psi_S$ and $\beta_R=\pi-\Psi_R$.  The two outer vertices contribute a net angle $\pi$ in the matching regime, while the oriented curvature-area term is taken over the exterior Gauss-Bonnet strip bounded by the particle trajectory and the outer closure.  With this orientation convention, Eq. \eqref{eq_GB_J_sec3} reduces to
\begin{equation}
-\iint_{D_J}K_J\,dS_J
+
\int_{C_\Gamma}\kappa_J\,d\ell_J
+
\Psi_S-\Psi_R
=
0 .
\label{eq_GB_reduced_sec3}
\end{equation}
Combining this relation with the finite-distance definition of $\alpha_J$ gives the exact Jacobi Gauss-Bonnet representation
\begin{equation}
\alpha_J
=
-\iint_{D_J}K_J\,dS_J
+
\int_{C_\Gamma}\kappa_J\,d\ell_J
+
\phi_{RS}.
\label{eq_alpha_GB_general_sec3}
\end{equation}
No circular orbit has been used.  The only assumptions are that the Jacobi metric is regular and Riemannian on $D_J$ and that the projected trajectory supplies a simple boundary segment.

For an asymptotically flat Jacobi geometry with Minkowski reference at fixed asymptotic speed, the outer closure satisfies
\begin{equation}
\int_{C_\Gamma}\kappa_J\,d\ell_J
\longrightarrow
-\phi_{RS},
\qquad
\alpha_J
=
-\iint_{D_J}K_J\,dS_J .
\label{eq_asymptotic_closure_sec3}
\end{equation}
For non-Euclidean references this cancellation is generally incomplete, and the closure term must be retained or combined consistently with the reference curvature.

The primitive reduction follows from the fact that any diagonal two-dimensional metric depending only on $r$ may be written as
\begin{align}
d\ell_J^2&=E_J(r)\,dr^2+G_J(r)\,d\phi^2,
\nn
\Gamma^\phi_{r\phi}(g_J)
&=
\frac{1}{2}\frac{d}{dr}\ln G_J(r).
\label{eq_generic_J_metric_sec3}
\end{align}
A direct curvature computation gives
\begin{align}
K_J
&=
-\frac{1}{\sqrt{\det g_J}}
\frac{d}{dr}
\left[
\frac{\sqrt{\det g_J}}{g^J_{rr}}
\Gamma^\phi_{r\phi}(g_J)
\right],
\nn
\mathcal{D}_J(r)&\equiv K_J(r)\sqrt{\det g_J(r)}
=
-\frac{dF_J}{dr}.
\label{eq_curvature_total_derivative_sec3}
\end{align}
For the Jacobi metric of Eq. \eqref{eq_jacobi_metric_sec2}, the curvature-density generator is
\begin{equation}
F_J(r)
=
\frac{1}{2}\sqrt{\frac{C(r)}{B(r)}}
\left[
\frac{C'(r)}{C(r)}
-
\frac{A'(r)}{A(r)}
-
\frac{A'(r)}{e^2-A(r)}
\right].
\label{eq_FJ_static_spherical_sec3}
\end{equation}
The final term is the explicit timelike correction.  It disappears only in the formal optical limit after the energy dependence of the Jacobi conformal factor becomes irrelevant.

Let $\mathcal{P}_J$ be a primitive of the radial curvature density.  Then
\begin{align}
\frac{d\mathcal{P}_J}{dr}&=\mathcal{D}_J(r),
\nn
\mathcal{P}_J(r)&=-F_J(r)+\mathcal{C}_J,
\nn
\iint_{D_J}K_J\,dS_J
&=
\int_{\phi_S}^{\phi_R}
\left[
\mathcal{P}_J(r_\Gamma)-\mathcal{P}_J(r(\phi))
\right]d\phi .
\label{eq_primitive_strip_sec3}
\end{align}
The additive constant $\mathcal{C}_J$ cancels from the exact strip integral.  The problem is therefore not the existence of a primitive, but the choice of a useful representative when the closure is moved to a reference end or when a weak-field reference trajectory is used.

\subsection{Reference-renormalized primitive}
\label{subsec_reference_renormalized_primitive}
We introduce a reference static spherical spacetime and its Jacobi metric at the same conserved energy $e$:
\begin{align}
ds_{\rm ref}^2
&=
-A_{\rm ref}(r)dt^2+B_{\rm ref}(r)dr^2+C_{\rm ref}(r)d\Omega^2,
\nn
d\ell_{J,{\rm ref}}^2
&=
\left[e^2-A_{\rm ref}(r)\right]
\left[
\frac{B_{\rm ref}(r)}{A_{\rm ref}(r)}dr^2
+
\frac{C_{\rm ref}(r)}{A_{\rm ref}(r)}d\phi^2
\right].
\label{eq_reference_metric_jacobi_sec3}
\end{align}
The fixed-energy condition is essential.  Here \textit{the same $e$} means that the energy is evaluated with respect to static Killing generators whose relative normalization has already been fixed.  Under a common constant rescaling $t'=\kappa t$, the metric coefficient and conserved energy transform as $A'=A/\kappa^2$ and $e'=e/\kappa$.  The Jacobi metric itself is unchanged because
\begin{align}
&\left(e'^2-A'\right)
\left[
\frac{B}{A'}dr^2+\frac{C}{A'}d\phi^2
\right]
\nn &=
\left(e^2-A\right)
\left[
\frac{B}{A}dr^2+\frac{C}{A}d\phi^2
\right].
\label{eq_common_time_rescaling_jacobi_sec3}
\end{align}
Thus a common rescaling changes only the shared energy label, whereas independent rescalings make equality of the numerical labels ambiguous.  We therefore fix the relative normalization of the physical and reference Killing generators before performing the reference subtraction and keep that choice fixed thereafter.

If the physical and reference metrics are written with different radial labels, the subtraction is performed on corresponding symmetry two-spheres.  Equivalently, we use the areal radius as the geometric matching variable,
\begin{equation}
\rho=\sqrt{C(r)}=\sqrt{C_{\rm ref}(r_{\rm ref})}.
\label{eq_areal_matching_sec3}
\end{equation}
In areal gauge this simply identifies the same coordinate value $r$ in the physical and reference functions.

The reference curvature density and the curvature-density discrepancy are defined by
\begin{align}
\mathcal{D}_{J,{\rm ref}}(r)
&\equiv
K_{J,{\rm ref}}(r)\sqrt{\det g_{J,{\rm ref}}(r)},
\nn
\Delta\mathcal{D}_J(r)
&\equiv
\mathcal{D}_J(r)-\mathcal{D}_{J,{\rm ref}}(r).
\label{eq_reference_discrepancy_density_sec3}
\end{align}
Let $r_\star$ denote the matching end.  It may be spatial infinity for an asymptotically flat comparison, a large matching sphere, or a finite endpoint inside a static patch.  We assume
\begin{equation}
\int_r^{r_\star}
\left|
\Delta\mathcal{D}_J(u)
\right|du
<\infty
\label{eq_integrability_sec3}
\end{equation}
for every $r$ in the scattering domain. This condition requires only the accumulated curvature-density discrepancy to be integrable; it does not require the physical and reference metrics to coincide pointwise.

The reference-renormalized Jacobi primitive is
\begin{align}
\mathcal{P}^{\rm ren}_J(r)
&\equiv
-\int_r^{r_\star}\Delta\mathcal{D}_J(u)\,du,
\nn
\frac{d\mathcal{P}^{\rm ren}_J}{dr}
&=
\Delta\mathcal{D}_J(r),
\nn
\mathcal{P}^{\rm ren}_J(r_\star)&=0 .
\label{eq_renormalized_primitive_sec3}
\end{align}
Thus the additive freedom is fixed by endpoint matching rather than by an auxiliary circular orbit.

When the reference density is itself integrable toward the same end, we also define
\begin{align}
\mathcal{P}^{\star}_{J,{\rm ref}}(r)
&\equiv
-\int_r^{r_\star}
\mathcal{D}_{J,{\rm ref}}(u)\,du,
\nn
\frac{d\mathcal{P}^{\star}_{J,{\rm ref}}}{dr}
&=
\mathcal{D}_{J,{\rm ref}}(r).
\label{eq_reference_star_primitive_sec3}
\end{align}
This is a convenience, not an additional assumption on the physical geometry.  If the reference primitive is not separately useful, the reference contribution should be kept in boundary form.

Using $\mathcal{D}_J=\mathcal{D}_{J,{\rm ref}}+\Delta\mathcal{D}_J$ in Eq. \eqref{eq_primitive_strip_sec3}, taking $r_\Gamma$ to the matching end, and then substituting into Eq. \eqref{eq_alpha_GB_general_sec3}, we obtain
\begin{align}
\alpha_J
&=
\int_{\phi_S}^{\phi_R}
\left[
\mathcal{P}^{\star}_{J,{\rm ref}}(r(\phi))
+
\mathcal{P}^{\rm ren}_J(r(\phi))
\right]d\phi
\nn &+
\int_{C_\Gamma}\kappa_J\,d\ell_J
+
\phi_{RS}.
\label{eq_master_formula_ref_primitive_sec3}
\end{align}
This is the reference-renormalized Jacobi master formula when the reference primitive is admissible.

The equivalent boundary-reference form is
\begin{align}
&\alpha_J
=
-\int_{\phi_S}^{\phi_R}
\left[
\int_{r(\phi)}^{r_\Gamma}
\mathcal{D}_{J,{\rm ref}}(r)\,dr
\right]d\phi
\nn &+
\int_{\phi_S}^{\phi_R}
\mathcal{P}^{\rm ren}_J(r(\phi))\,d\phi
+
\int_{C_\Gamma}\kappa_J\,d\ell_J
+
\phi_{RS}.
\label{eq_master_formula_boundary_ref_sec3}
\end{align}
This form is preferable for finite static patches and non-Euclidean references, because the pure-reference curvature and closure terms remain explicitly paired.

For an asymptotically flat spacetime with Minkowski Jacobi reference at the same asymptotic speed, $\mathcal{D}_{J,{\rm ref}}=0$ and the closure contribution cancels $\phi_{RS}$.  The master formula reduces to
\begin{equation}
\alpha_J
=
\int_{\phi_S}^{\phi_R}
\mathcal{P}^{\rm ren}_J(r(\phi))\,d\phi .
\label{eq_master_asymptotically_flat_sec3}
\end{equation}
If the physical Jacobi geometry coincides with the reference Jacobi geometry, then $\Delta\mathcal{D}_J=0$, so $\mathcal{P}^{\rm ren}_J=0$ and the relative bending vanishes by construction.

\subsection{Additive-gauge equivalence and validity}
\label{subsec_additive_gauge_equivalence_validity}
\noindent\textbf{Proposition.} Let $I$ be a connected radial interval on which the Jacobi geometry is regular, and let $\mathcal{P}_{J,1}$ and $\mathcal{P}_{J,2}$ be two $C^1$ primitives of the same curvature density $\mathcal{D}_J$.  If $d\mathcal{P}_{J,1}/dr=d\mathcal{P}_{J,2}/dr=\mathcal{D}_J(r)$ for $r\in I$, then $\mathcal{P}_{J,1}-\mathcal{P}_{J,2}$ is constant on $I$.  The exact curvature strip is therefore unchanged under the shift $\mathcal{P}_J\mapsto\mathcal{P}_J+\mathcal{C}$ with constant $\mathcal{C}$.  In particular, whenever the reference end and an admissible circular Jacobi orbit belong to the same connected regular interval, endpoint normalization and circular-orbit normalization give the same exact curvature contribution.

\noindent\textit{Proof.} Subtracting the two primitive relations gives $d(\mathcal{P}_{J,1}-\mathcal{P}_{J,2})/dr=0$.  Connectedness of $I$ then makes the difference constant.  The strip formula in Eq. \eqref{eq_primitive_strip_sec3} contains only the difference between the primitive evaluated at the outer end and along the trajectory, so the same constant cancels identically.  The endpoint-normalized and circular-orbit-normalized representatives considered below are therefore members of this same additive class.\hfill$\square$

The circular-orbit construction is recovered as a special choice of primitive representative.  A constant-radius Jacobi curve is geodesic when $G_J'(r_{\rm co})=0$, where $G_J=C(e^2-A)/A$.  Since $F_J$ is proportional to $\Gamma^\phi_{r\phi}=G_J'/(2G_J)$, an admissible circular orbit satisfies
\begin{align}
\left.
\frac{d}{dr}
\left[
\frac{C(r)
\left(e^2-A(r)\right)}{A(r)}
\right]\right|_{r=r_{\rm co}}
=0,
\nn
F_J(r_{\rm co})=0,
\qquad
\mathcal{P}^{\rm orb}_J(r)\equiv -F_J(r).
\label{eq_circular_orbit_primitive_sec3}
\end{align}
The use of the circular-orbit condition as a primitive normalization follows the Jacobi/optical circular-orbit construction of Ref.~\cite{Li:2020wvn}. Thus the orbit-normalized primitive is simply one representative of the same primitive class.

Let $\mathcal{P}^{\star}_J$ be the physical primitive normalized at the reference end, and let $\mathcal{P}^{\rm orb}_J$ be the primitive normalized at an admissible circular orbit.  Since both differentiate to $\mathcal{D}_J$, their difference is constant:
\begin{align}
\frac{d\mathcal{P}^{\star}_J}{dr}
&=
\frac{d\mathcal{P}^{\rm orb}_J}{dr}
=
\mathcal{D}_J(r),
\nn
\mathcal{P}^{\star}_J(r)
&=
\mathcal{P}^{\rm orb}_J(r)
-
\mathcal{P}^{\rm orb}_J(r_\star).
\label{eq_Pstar_Porb_relation_sec3}
\end{align}
Consequently, the curvature strip computed with the endpoint-normalized primitive is identical to the circular-orbit-normalized expression:
\begin{align}
&-\int_{\phi_S}^{\phi_R}
\mathcal{P}^{\star}_J(r(\phi))\,d\phi
\nn & =\int_{\phi_S}^{\phi_R}
\left[
\mathcal{P}^{\rm orb}_J(r_\star)
-
\mathcal{P}^{\rm orb}_J(r(\phi))
\right]d\phi .
\label{eq_strip_equivalence_sec3}
\end{align}
Equation \eqref{eq_strip_equivalence_sec3} is the circular-orbit specialization of the proposition above.  Whenever both normalizations are admissible, they give the same exact curvature contribution.  When the circular orbit is absent or outside the allowed Jacobi domain, the reference-normalized primitive remains available provided the discrepancy condition \eqref{eq_integrability_sec3} holds.

The primitive construction is also robust under monotonic radial reparameterizations.  If $r\mapsto\bar r$ is used only as a radial relabeling of the same symmetry two-spheres, then
\begin{align}
\bar{\mathcal{D}}_J(\bar r)\,d\bar r
&=
\mathcal{D}_J(r)\,dr,
\nn
\Delta\bar{\mathcal{D}}_J(\bar r)\,d\bar r
&=
\Delta\mathcal{D}_J(r)\,dr,
\nn
\bar{\mathcal{P}}^{\rm ren}_J(\bar r)
&=
\mathcal{P}^{\rm ren}_J(r).
\label{eq_coordinate_robustness_sec3}
\end{align}
Thus the one-dimensional primitive notation is coordinate-dependent, but the endpoint-normalized primitive is a scalar function on the radial orbit space once the reference matching is specified geometrically.

The domain of validity is now clear.  The trajectory and the Gauss-Bonnet domain must remain in the Riemannian Jacobi region of Eq. \eqref{eq_jacobi_domain_sec2}; the endpoint inequalities in Eq. \eqref{eq_endpoint_admissibility_sec2} must hold; the projected path must be single-pass with one turning point; and the discrepancy density must satisfy Eq. \eqref{eq_integrability_sec3}.  Bound motion, multiple winding, horizon crossing, and stationary Randers-Finsler Jacobi geometries require separate boundary analyses and are not included in the present construction.

\section{Weak-deflection expansion protocol}
\label{sec_weak_expansion_protocol}
The master formulas in Sec. \ref{sec_jacobi_reference_renormalization} are exact within the assumed Jacobi domain.  To obtain explicit finite-distance bending angles, one must choose a reference geometry, expand the orbit and endpoint quantities relative to that geometry, and evaluate the reference-renormalized primitive consistently at the same order.  This section gives the expansion rules used in the bumblebee, Kottler, and JNW applications.

\subsection{Reference trajectory and turning-point expansion}
\label{subsec_reference_trajectory_turning_expansion}
It is useful to write the radial branch factor in a compact form.  With $H(r,e,\ell)$ defined in Eq. \eqref{eq_radial_equation_H_sec2}, we set
\begin{align}
\mathcal{R}(r,e,\ell)
&\equiv
\frac{\ell\sqrt{A(r)B(r)}}{\sqrt{C(r)}\sqrt{H(r,e,\ell)}} ,
\nn
\frac{d\phi}{dr}&=\pm \mathcal{R}(r,e,\ell).
\label{eq_R_factor_sec4}
\end{align}
For any scalar function $Q(r)$ evaluated along a single-turning-point orbit, the angular integral can be replaced by the two radial branches,
\begin{align}
&\int_{\phi_S}^{\phi_R}Q(r(\phi))\,d\phi
\nn &=  \int_{r_0}^{r_R}Q(r)\mathcal{R}(r,e,\ell)\,dr
+
\int_{r_0}^{r_S}Q(r)\mathcal{R}(r,e,\ell)\,dr .
\label{eq_angular_to_radial_sec4}
\end{align}
This identity is exact before any weak-field expansion.

Let the reference geometry be described by functions $A_0(r)$, $B_0(r)$, and $C_0(r)$ at the same conserved energy $e$.  Its orbit function and turning point are
\begin{equation}
H_0(r,e,\ell)
=
C_0(r)\left[e^2-A_0(r)\right]-A_0(r)\ell^2,
\qquad
H_0(r_0^{(0)},e,\ell)=0 .
\label{eq_H0_turning_sec4}
\end{equation}
The reference angular span is therefore
\begin{equation}
\phi_{RS}^{(0)}
=
\int_{r_0^{(0)}}^{r_R}\mathcal{R}_0(r,e,\ell)\,dr
+
\int_{r_0^{(0)}}^{r_S}\mathcal{R}_0(r,e,\ell)\,dr ,
\label{eq_reference_phi_span_sec4}
\end{equation}
where $\mathcal{R}_0$ is obtained from Eq. \eqref{eq_R_factor_sec4} by replacing the physical metric functions by the reference functions.

The reference endpoint angle is controlled by
\begin{equation}
\chi_0(r,e,\ell)
\equiv
\ell
\sqrt{
\frac{A_0(r)}
{C_0(r)\left[e^2-A_0(r)\right]}
}.
\label{eq_reference_chi_sec4}
\end{equation}
For the Euclidean Minkowski Jacobi reference, the corresponding no-lens identity is
\begin{equation}
\arcsin\chi_0(r_R)
+
\arcsin\chi_0(r_S)
-
\pi
+
\phi_{RS}^{(0)}
=
0 .
\label{eq_reference_no_lens_sec4}
\end{equation}
Equation \eqref{eq_reference_no_lens_sec4} should be understood as a Euclidean-reference identity.  For conical, cosmological, or otherwise non-Euclidean references, the reference curvature and the oriented closure term must be retained explicitly; in such cases the reference contribution need not vanish when written in the original angular coordinate.

We introduce a formal bookkeeping parameter $\varepsilon$ and expand the physical metric about the reference metric,
\begin{align}
A&=A_0+\varepsilon A_1+\varepsilon^2A_2+\cdots,
\nn
B&=B_0+\varepsilon B_1+\varepsilon^2B_2+\cdots,
\nn
C&=C_0+\varepsilon C_1+\varepsilon^2C_2+\cdots .
\label{eq_metric_expansion_sec4}
\end{align}
The parameter $\varepsilon$ is set to unity after the desired order is extracted.  In applications it counts the localized mass scale, Lorentz-violating deformation, scalar-charge contribution, or any chosen combination of weak-field parameters.

The turning point is expanded as
\begin{align}
r_0
&=
r_0^{(0)}+\varepsilon r_0^{(1)}+\varepsilon^2r_0^{(2)}+\cdots ,
\nn
H&=H_0+\varepsilon H_1+\varepsilon^2H_2+\cdots .
\label{eq_turning_H_expansion_sec4}
\end{align}
Expanding $H(r_0,e,\ell)=0$ gives the first correction
\begin{align}
r_0^{(1)}
&=
-\frac{H_1(r_0^{(0)},e,\ell)}
{H_0'(r_0^{(0)},e,\ell)},
\nn
H_1
&=
C_1(e^2-A_0)-C_0A_1-A_1\ell^2 .
\label{eq_first_turning_shift_sec4}
\end{align}
This shift must be retained whenever the angular-span integral is expanded directly, because $\mathcal{R}$ has a square-root singularity at the lower limit.  A safe alternative is to change variables so that the lower integration limit is fixed before expanding.

\subsection{Primitive expansion from the generator}
\label{subsec_primitive_expansion_generator}
The most economical way to compute the reference-renormalized primitive is to use the curvature-density generator.  For any static spherical Jacobi metric,
\begin{align}
F_J[A,B,C]
&=
\frac{1}{2}\sqrt{\frac{C}{B}}
\left[
\frac{C'}{C}
-
\frac{A'}{A}
-
\frac{A'}{e^2-A}
\right],
\nn
\mathcal{D}_J&=-F_J' .
\label{eq_FJ_functional_sec4}
\end{align}
Let $F_{J,0}=F_J[A_0,B_0,C_0]$ and $\Delta F_J=F_J-F_{J,0}$.  Since $\Delta\mathcal{D}_J=-\Delta F_J'$, the renormalized primitive is
\begin{equation}
\mathcal{P}^{\rm ren}_J(r)
=
\Delta F_J(r_\star)-\Delta F_J(r).
\label{eq_Pren_DeltaF_sec4}
\end{equation}
Thus the endpoint subtraction is implemented algebraically once $F_J$ is known.

Expanding the generator as
\begin{align}
F_J
&=
F_J^{(0)}+\varepsilon F_J^{(1)}+\varepsilon^2F_J^{(2)}+\cdots,
\nn
F_J^{(0)}&=F_{J,0},
\label{eq_FJ_expansion_sec4}
\end{align}
we define
\begin{equation}
\mathcal{T}_0
\equiv
\frac{C_0'}{C_0}
-
\frac{A_0'}{A_0}
-
\frac{A_0'}{e^2-A_0}.
\label{eq_T0_sec4}
\end{equation}
The first-order generator correction is
\begin{equation}
F_J^{(1)}
=
\frac{1}{2}\sqrt{\frac{C_0}{B_0}}
\left[
\mathcal{T}_1
+
\frac{1}{2}
\left(
\frac{C_1}{C_0}
-
\frac{B_1}{B_0}
\right)\mathcal{T}_0
\right],
\label{eq_FJ1_sec4}
\end{equation}
where
\begin{equation}
\mathcal{T}_1
=
\frac{C_1'}{C_0}
-
\frac{C_0'C_1}{C_0^2}
-
\frac{A_1'}{A_0}
+
\frac{A_0'A_1}{A_0^2}
-
\frac{A_1'}{e^2-A_0}
-
\frac{A_0'A_1}{(e^2-A_0)^2}.
\label{eq_T1_sec4}
\end{equation}
At order $\varepsilon^n$, Eq. \eqref{eq_Pren_DeltaF_sec4} gives
\begin{equation}
\mathcal{P}^{(n)}_J(r)
=
F_J^{(n)}(r_\star)-F_J^{(n)}(r).
\label{eq_Pn_from_Fn_sec4}
\end{equation}
For an infinite matching end, the first term often vanishes.  For finite static patches it generally does not, and it must be kept.

In an asymptotically flat areal gauge, the Minkowski reference is
\begin{equation}
A_0=1,
\qquad
B_0=1,
\qquad
C_0=r^2 .
\label{eq_minkowski_reference_sec4}
\end{equation}
If the areal coordinate is unchanged at first order, so that $C_1=0$, then Eqs. \eqref{eq_FJ1_sec4}-\eqref{eq_T1_sec4} reduce to
\begin{equation}
F_J^{(1)}
=
-\frac{re^2}{2(e^2-1)}A_1'
-\frac{1}{2}B_1
=
-\frac{r}{2v^2}A_1'
-\frac{1}{2}B_1 .
\label{eq_FJ1_asymptotic_sec4}
\end{equation}
When $F_J^{(1)}(\infty)=0$, the leading asymptotically flat primitive is therefore
\begin{equation}
\mathcal{P}^{(1)}_J(r)
=
\frac{r}{2v^2}A_1'(r)
+
\frac{1}{2}B_1(r).
\label{eq_PJ1_asymptotic_sec4}
\end{equation}
This shortcut will be useful for asymptotically flat sectors.  It should not be applied to conical or de Sitter references without retaining the appropriate closure and reference-curvature terms.

\subsection{Application rules and consistency checks}
\label{subsec_application_rules_checks}
The practical first-order rule is the following.  If the reference curvature and closure terms already give the appropriate reference identity, then the localized bending contribution can be evaluated by inserting the first nonzero discrepancy primitive on the reference trajectory.  For a Euclidean Minkowski reference this identity reduces to Eq. \eqref{eq_reference_no_lens_sec4}; for conical or de Sitter references it is instead the full reference Gauss-Bonnet identity, including the reference curvature and the oriented closure term.  Equivalently, one may use Eq. \eqref{eq_angular_to_radial_sec4} with $Q=\mathcal{P}_J^{(1)}$ and $\mathcal{R}=\mathcal{R}_0$.  Corrections to the path multiply an already first-order primitive and therefore enter only at second order.

For conical references, the reference geometry is flat but the closure term is not Euclidean.  The angular range of the reference trajectory and the asymptotic value of $\kappa_Jd\ell_J$ must therefore be kept explicitly.  This is the relevant organization for the Schwarzschild-like bumblebee spacetime.

For Kottler-type geometries, the perturbation is the localized mass parameter, while the cosmological term belongs to the de Sitter reference.  The reference part of the master formula supplies the no-localized-mass identity inside the static patch.  The mass-induced bending is then obtained from the de Sitter discrepancy primitive, together with the endpoint and orbit expansions performed at fixed conserved energy.

For naked-singularity examples such as the Janis-Newman-Winicour spacetime, the asymptotically flat reference is again the Minkowski Jacobi metric.  The circular-orbit condition is not used.  The calculation instead asks whether the discrepancy primitive is well defined and whether the resulting finite-distance bending has the correct Schwarzschild and ultra-relativistic limits.

The following checks will be imposed in the applications. First, the discrepancy contribution must vanish when the physical Jacobi geometry is set equal to the chosen reference geometry.  The operational finite-distance angle may still contain a pure-reference contribution in non-Euclidean reference geometries, such as conical or de Sitter backgrounds.  Second, the $\lambda\to1$, $\Lambda\to0$, or $\gamma\to1$ limits must reproduce the appropriate Schwarzschild massive-particle expression when the corresponding metric reduces to Schwarzschild.  Third, the ultra-relativistic limit $e\to\infty$ must reproduce the optical finite-distance result.  In asymptotically flat sectors this is equivalently the limit $v\to1$ of Eq. \eqref{eq_v_b_def_sec2}. Fourth, whenever an admissible circular timelike orbit exists, the reference-renormalized primitive must agree with the circular-orbit-normalized prescription up to a constant that cancels from the exact curvature strip.

\section{Schwarzschild-like bumblebee spacetime}
\label{sec_bumblebee_application}
We first apply the reference-renormalized Jacobi primitive to a spacetime whose no-mass limit is not Euclidean but conical.  This example is useful because it tests two features at once: the primitive must reproduce the circular-orbit Jacobi result in the localized Schwarzschild sector, while the finite-distance formula must also retain the closure contribution associated with the conical reference geometry.  We use the Schwarzschild-like bumblebee metric considered in the circular-orbit Jacobi analysis of Li, Zhang, and \"Ovg\"un \cite{Li:2020wvn}.

\subsection{Conical reference geometry and primitive}
\label{subsec_bumblebee_reference_primitive}
The spacetime is described by (previously used in \cite{Li:2020dln,Li:2020wvn})
\begin{align}
ds^2
&=
-\left(1-\frac{2M}{r}\right)dt^2
+
\frac{\lambda^2}{1-\frac{2M}{r}}dr^2
+
r^2d\Omega^2,
\nn
\lambda&=\sqrt{1+l},
\label{eq_bumblebee_metric_sec5}
\end{align}
with $\lambda>0$. Thus
\begin{align}
A(r)&=1-\frac{2M}{r},
\nn
B(r)&=\frac{\lambda^2}{1-\frac{2M}{r}},
\nn
C(r)&=r^2 .
\label{eq_bumblebee_functions_sec5}
\end{align}
The massless member is not the Euclidean spatial reference unless $\lambda=1$.  It is instead the conical reference
\begin{equation}
ds_0^2=-dt^2+\lambda^2dr^2+r^2d\Omega^2 .
\label{eq_bumblebee_reference_metric_sec5}
\end{equation}
At fixed conserved energy $e$, the reference Jacobi metric is
\begin{equation}
d\ell_{J,0}^2
=
(e^2-1)\left(\lambda^2dr^2+r^2d\phi^2\right).
\label{eq_bumblebee_reference_jacobi_sec5}
\end{equation}
It is flat as a two-dimensional metric after the angular rescaling $\phi\mapsto\phi/\lambda$, but it is conical relative to the original rotational coordinate $\phi$.  Therefore the outer closure term cannot be replaced by the Euclidean cancellation used in asymptotically flat Minkowski references.

The Jacobi curvature-density generator for Eq. \eqref{eq_bumblebee_metric_sec5} follows directly from Eq. \eqref{eq_FJ_functional_sec4}.  Since the bumblebee deformation enters only through $B(r)$, it rescales the Schwarzschild generator by $1/\lambda$:
\begin{align}
F^{\rm B}_J(r)
&=
\frac{1}{\lambda}
\Bigg[
\sqrt{1-\frac{2M}{r}}
-
\frac{M}{r\sqrt{1-\frac{2M}{r}}}
\nn &-
\frac{M\sqrt{1-\frac{2M}{r}}}
{r\left(e^2-1+\frac{2M}{r}\right)}
\Bigg].
\label{eq_FJ_bumblebee_exact_sec5}
\end{align}
For the conical reference, $F_{J,0}=1/\lambda$.  The reference-renormalized primitive is therefore
\begin{equation}
\mathcal{P}^{\rm ren}_{J,{\rm B}}(r)
=
\frac{1}{\lambda}-F^{\rm B}_J(r).
\label{eq_PJ_bumblebee_exact_sec5}
\end{equation}
Expanding for $M/r\ll1$ and using $e^2=(1-v^2)^{-1}$ gives
\begin{equation}
\mathcal{P}^{\rm ren}_{J,{\rm B}}(r)
=
\frac{M}{\lambda r}
\left(1+\frac{1}{v^2}\right)
+
\mathcal{O}\!\left(\frac{M^2}{r^2}\right).
\label{eq_PJ_bumblebee_leading_sec5}
\end{equation}
This primitive is the localized mass contribution normalized with respect to the conical no-mass background.  The pure conical part is not hidden in the primitive; it appears through the reference closure and the reference angular span.

\subsection{Finite-distance bending angle}
\label{subsec_bumblebee_finite_distance_angle}
The reference trajectory is obtained from Eq. \eqref{eq_R_factor_sec4} with the metric functions of Eq. \eqref{eq_bumblebee_reference_metric_sec5}.  With $b=\ell/\sqrt{e^2-1}$, one finds
\begin{align}
r^{(0)}(\phi)
&=
\frac{b}{\cos(\phi/\lambda)},
\nn
\phi_R^{(0)}
&=
\lambda\arccos\!\left(\frac{b}{r_R}\right),
\nn
\phi_S^{(0)}
&=
-\lambda\arccos\!\left(\frac{b}{r_S}\right).
\label{eq_bumblebee_reference_trajectory_sec5}
\end{align}
We define
\begin{align}
s_i&\equiv
\sqrt{1-\frac{b^2}{r_i^2}},
\nn
x_i&\equiv\frac{b^2}{r_i^2},
\nn
\Phi_0
&\equiv
\pi-\arcsin\!\left(\frac{b}{r_R}\right)
-\arcsin\!\left(\frac{b}{r_S}\right),
\label{eq_bumblebee_endpoint_symbols_sec5}
\end{align}
with $i = R,S$. Thus the reference angular span is $\phi_{RS}^{(0)}=\lambda\Phi_0$.  The local endpoint angles of the $M=0$ reference are the same functions of $b/r_i$ as in the Euclidean case, because $A_0=1$ and $C_0=r^2$.  Therefore the Euclidean no-lens identity is recovered only when $\lambda=1$; for $\lambda\neq1$, the oriented conical closure leaves the finite reference contribution $(\lambda-1)\Phi_0$ to the bending angle.

At first order in $M$, the primitive may be evaluated on the reference trajectory.  Using Eq. \eqref{eq_PJ_bumblebee_leading_sec5} and Eq. \eqref{eq_bumblebee_reference_trajectory_sec5},
\begin{align}
&\int_{\phi_S^{(0)}}^{\phi_R^{(0)}}
\mathcal{P}^{\rm ren}_{J,{\rm B}}(r^{(0)}(\phi))\,d\phi
\nn &=
\frac{M}{b}
\left(1+\frac{1}{v^2}\right)
(s_R+s_S)
+
\mathcal{O}\!\left(\frac{M^2}{b^2}\right).
\label{eq_bumblebee_primitive_integral_sec5}
\end{align}
At large radius, the conical closure has the local asymptotic form $\kappa_Jd\ell_J\to d\phi/\lambda$.  With the boundary orientation of the Gauss-Bonnet strip, however, its contribution is
\begin{equation}
\int_{C_\Gamma}\kappa_J\,d\ell_J
\longrightarrow
-\frac{\phi_{RS}}{\lambda}.
\label{eq_bumblebee_oriented_closure_sec5}
\end{equation}
Therefore the finite-distance Gauss-Bonnet identity takes the form
\begin{equation}
\alpha_{J,{\rm B}}
=
\int_{\phi_S}^{\phi_R}
\mathcal{P}^{\rm ren}_{J,{\rm B}}(r(\phi))\,d\phi
+
\left(1-\frac{1}{\lambda}\right)\phi_{RS}.
\label{eq_bumblebee_alpha_structure_sec5}
\end{equation}
The second term is absent only for $\lambda=1$.  It is the finite-distance remnant of the conical reference geometry.

The physical angular span, expanded through first order in $M/b$, is
\begin{align}
\phi_{RS}
&=
\lambda\Phi_0
+
\frac{M\lambda}{b}
\left[
s_R+s_S
+
\frac{1}{v^2}
\left(
\frac{1}{s_R}+\frac{1}{s_S}
\right)
\right]
\nn &+
\mathcal{O}\!\left(\frac{M^2}{b^2}\right).
\label{eq_bumblebee_phiRS_expansion_sec5}
\end{align}
The terms proportional to $1/s_i$ arise from the finite endpoint correction to the radial branch integrals.  They must be kept because the conical closure multiplies the full coordinate span.

Substituting Eqs. \eqref{eq_bumblebee_primitive_integral_sec5} and \eqref{eq_bumblebee_phiRS_expansion_sec5} into Eq. \eqref{eq_bumblebee_alpha_structure_sec5}, we obtain
\begin{align}
\alpha_{J,{\rm B}}
&=
(\lambda-1)\Phi_0
\nn &+
\frac{M}{bv^2}
\sum_{i=R,S}
\frac{
(1+v^2)\lambda
-
x_i(1+\lambda v^2)
}
{\sqrt{1-x_i}}
\nn &+
\mathcal{O}\!\left(\frac{M^2}{b^2}\right).
\label{eq_bumblebee_final_compact_sec5}
\end{align}
Equivalently, written without $x_i$,
\begin{align}
&\alpha_{J,{\rm B}}
=
(\lambda-1)
\left[
\pi-\arcsin\!\left(\frac{b}{r_R}\right)
-\arcsin\!\left(\frac{b}{r_S}\right)
\right]
\nn &+
\frac{M}{bv^2}
\sum_{i=R,S}
\frac{
(1+v^2)\lambda
-
\frac{b^2}{r_i^2}(1+\lambda v^2)
}
{\sqrt{1-\frac{b^2}{r_i^2}}}
+
\mathcal{O}\!\left(\frac{M^2}{b^2}\right).
\label{eq_bumblebee_final_explicit_sec5}
\end{align}
This is the finite-distance massive-particle bending angle in the Schwarzschild-like bumblebee spacetime through first order in the localized mass parameter.  The first term is purely conical; the second term is the finite-distance mass correction modified by both the particle speed and the conical deformation.

\subsection{Checks and comparison with circular-orbit normalization}
\label{subsec_bumblebee_checks_comparison}
Several checks are immediate.  First, setting $\lambda=1$ removes the conical reference contribution and gives the standard finite-distance Schwarzschild massive-particle result,
\begin{equation}
\alpha_{J,{\rm B}}\big|_{\lambda=1}
=
\frac{M}{b}
\left(1+\frac{1}{v^2}\right)
(s_R+s_S)
+
\mathcal{O}\!\left(\frac{M^2}{b^2}\right).
\label{eq_bumblebee_schwarzschild_limit_sec5}
\end{equation}
Thus the Schwarzschild case is recovered as a limit of the bumblebee calculation rather than as a separate application.

Second, taking the infinite-distance limit $r_R,r_S\to\infty$ gives
\begin{equation}
\alpha_{J,{\rm B}}^{\infty}
=
(\lambda-1)\pi
+
\frac{2M\lambda}{b}
\left(1+\frac{1}{v^2}\right)
+
\mathcal{O}\!\left(\frac{M^2}{b^2}\right).
\label{eq_bumblebee_infinite_distance_sec5}
\end{equation}
The first term is the conical asymptotic contribution, while the second term reduces to the usual velocity-dependent Schwarzschild bending when $\lambda=1$.

Third, the ultra-relativistic limit gives the finite-distance optical result for the same conical background:
\begin{align}
\alpha_{{\rm opt},{\rm B}}
&=
(\lambda-1)\Phi_0
+
\frac{M}{b}
\sum_{i=R,S}
\frac{
2\lambda
-
x_i(1+\lambda)
}
{\sqrt{1-x_i}}
\nn &+
\mathcal{O}\!\left(\frac{M^2}{b^2}\right).
\label{eq_bumblebee_null_limit_sec5}
\end{align}
For $\lambda=1$ and $r_R,r_S\to\infty$, this further reduces to $4M/b$.

Finally, the relation with the circular-orbit primitive is transparent.  The Jacobi circular-orbit condition in the bumblebee metric is independent of the constant radial rescaling $\lambda$ and gives
\begin{align}
e^2(r_{\rm co}-3M)
&=
r_{\rm co}
\left(1-\frac{2M}{r_{\rm co}}\right)^2,
\nn
r_{\rm co}
&=
\frac{M}{2v^2}
\left[
4v^2-1+\sqrt{1+8v^2}
\right].
\label{eq_bumblebee_circular_orbit_sec5}
\end{align}
At this radius $F^{\rm B}_J(r_{\rm co})=0$, so the orbit-normalized primitive is $\mathcal{P}^{\rm orb}_{J,{\rm B}}=-F^{\rm B}_J$.  Comparing with Eq. \eqref{eq_PJ_bumblebee_exact_sec5},
\begin{equation}
\mathcal{P}^{\rm ren}_{J,{\rm B}}(r)
=
\mathcal{P}^{\rm orb}_{J,{\rm B}}(r)
+
\frac{1}{\lambda}.
\label{eq_bumblebee_primitive_constant_shift_sec5}
\end{equation}
The two prescriptions therefore differ only by a constant, exactly as required by the additive-gauge proposition of Sec. \ref{subsec_additive_gauge_equivalence_validity}.  The constant has no effect on the exact curvature strip when the closure and reference terms are treated consistently.  This confirms that the reference-renormalized calculation reproduces the circular-orbit Jacobi result in the common domain, while making explicit the conical reference contribution that must accompany finite-distance bending in the bumblebee geometry.

\section{Kottler spacetime with de Sitter Jacobi reference}
\label{sec_kottler_application}
We next consider a case in which the reference geometry is intrinsically non-Euclidean.  For the Kottler spacetime, the localized lens parameter should be separated from the cosmological background.  The appropriate comparison geometry is therefore not Minkowski space, but the de Sitter member obtained by setting the localized mass parameter to zero while keeping the same conserved particle energy.  This section formulates the finite-distance expansion in that reference geometry and identifies the mass-induced discrepancy primitive.

\subsection{Fixed-energy de Sitter reference}
\label{subsec_kottler_desitter_reference}
We write the Kottler metric as
\begin{align}
ds^2
&=
-A(r)dt^2+A(r)^{-1}dr^2+r^2d\Omega^2,
\nn
A(r)&=1-\frac{r_g}{r}-\frac{\Lambda r^2}{3}.
\label{eq_kottler_metric_sec6}
\end{align}
The same Kottler or Schwarzschild-de Sitter form is commonly used in finite-distance and massive-particle deflection analyses, although the present calculation organizes the mass contribution relative to a de Sitter Jacobi reference \cite{Takizawa:2020egm,He:2020eah}. The localized perturbation is $r_g$, while $\Lambda$ is treated as part of the reference background.  Thus the de Sitter reference is
\begin{align}
A_0(r)&=a(r)\equiv1-\frac{\Lambda r^2}{3},
\nn
B_0(r)&=a(r)^{-1},
\nn
C_0(r)&=r^2 .
\label{eq_desitter_reference_functions_sec6}
\end{align}
For the Kottler-de Sitter pair, we fix the relative normalization of the static Killing fields without appealing to spatial infinity.  We obtain the reference member by setting $r_g=0$ while retaining the same coordinate $t$.  Hence the generators $\xi_{\rm Kot}=\partial_t$ and $\xi_{\rm dS}=\partial_t$ are identified before the fixed-energy comparison is made.  We do not subsequently rescale $t$ in one geometry independently of the other.  The conserved label $e=-u_\mu\xi^\mu$ therefore denotes the same Killing-energy normalization on both sides of the subtraction.

At fixed conserved energy $e$, the reference Jacobi metric is
\begin{equation}
d\ell_{J,0}^2
=
\left[e^2-a(r)\right]
\left[
\frac{dr^2}{a(r)^2}
+
\frac{r^2}{a(r)}d\phi^2
\right].
\label{eq_desitter_jacobi_metric_sec6}
\end{equation}
The construction is restricted to the common static and Jacobi region
\begin{align}
&a(r)>0,
\nn
&e^2>a(r),
\nn
&0<r<\sqrt{\frac{3}{\Lambda}}
\quad
(\Lambda>0).
\label{eq_desitter_domain_sec6}
\end{align}
For $\Lambda=0$, this reference reduces smoothly to the Minkowski Jacobi geometry.

It is useful to introduce
\begin{equation}
W(r)\equiv e^2-a(r)=e^2-1+\frac{\Lambda r^2}{3}.
\label{eq_W_definition_sec6}
\end{equation}
The de Sitter reference orbit is governed by
\begin{align}
H_0(r,e,\ell)
&=
r^2W(r)-a(r)\ell^2,
\nn
\mathcal{R}_0(r,e,\ell)
&=
\frac{\ell}{r\sqrt{H_0(r,e,\ell)}} .
\label{eq_desitter_H_R_sec6}
\end{align}
The reference endpoint angle is
\begin{align}
\chi_0(r,e,\ell)
&=
\ell\sqrt{\frac{a(r)}{r^2W(r)}} ,
\nn
\Psi_R^{(0)}&=\arcsin\chi_0(r_R),
\nn
\Psi_S^{(0)}&=\pi-\arcsin\chi_0(r_S).
\label{eq_desitter_endpoint_angle_sec6}
\end{align}
The pure de Sitter reference supplies the no-localized-mass comparison geometry.  In the relative finite-distance prescription, the bending attributed to the localized mass is obtained by expanding about this reference at fixed $e$, not by discarding $\Lambda$.

\subsection{Mass-discrepancy primitive}
\label{subsec_kottler_mass_discrepancy_primitive}
We now expand the Kottler metric in the localized mass parameter,
\begin{align}
A(r)&=a(r)-\frac{r_g}{r},
\nn
B(r)&=A(r)^{-1},
\nn
F_J^{\rm Kot}(r)
&=
F_{J,0}^{\rm dS}(r)
+
r_g F_{J,M}^{\rm dS}(r)
+
\mathcal{O}(r_g^2).
\label{eq_kottler_F_expansion_sec6}
\end{align}
The reference generator is
\begin{equation}
F_{J,0}^{\rm dS}(r)
=
\sqrt{a(r)}
-
\frac{r\,a'(r)}{2\sqrt{a(r)}}
-
\frac{r\sqrt{a(r)}\,a'(r)}{2W(r)} .
\label{eq_FJ_desitter_reference_sec6}
\end{equation}
The first-order mass-discrepancy generator can be written compactly as
\begin{widetext}
\begin{align}
F_{J,M}^{\rm dS}(r)
=
\frac{
\left[1-3a(r)\right]W(r)^2
-a(r)W(r)
+2a(r)^2\left[a(r)-1\right]
}
{
2r\,a(r)^{3/2}W(r)^2
}.
\label{eq_FJM_desitter_compact_sec6}
\end{align}
Equivalently, in terms of $\Lambda$, this is
\begin{align}
F_{J,M}^{\rm dS}(r)
=
\frac{
\Lambda^3r^6
+18\Lambda^2e^2r^4
-9\Lambda^2r^4
+27\Lambda e^4r^2
-81\Lambda e^2r^2
+27\Lambda r^2
-54e^4
+81e^2
-27
}
{
6r\left(1-\frac{\Lambda r^2}{3}\right)^{3/2}
\left(\Lambda r^2+3e^2-3\right)^2
}.
\label{eq_FJM_desitter_explicit_sec6}
\end{align}
\end{widetext}
The denominator contains the static-patch factor and the Jacobi conformal factor.  Hence the expansion is meaningful only inside the region specified in Eq. \eqref{eq_desitter_domain_sec6}.

Let $r_\star$ be a matching sphere inside the de Sitter static patch.  The mass-induced reference-renormalized primitive is
\begin{equation}
\mathcal{P}^{\rm ren}_{J,{\rm Kot}}(r,r_\star)
=
r_g
\left[
F_{J,M}^{\rm dS}(r_\star)
-
F_{J,M}^{\rm dS}(r)
\right]
+
\mathcal{O}(r_g^2).
\label{eq_PJ_kottler_ren_sec6}
\end{equation}
Changing $r_\star$ shifts the primitive by a constant.  At the exact level this constant is canceled by the corresponding reference and closure terms in the Gauss-Bonnet identity.  At finite perturbative order, $r_\star$ should be chosen in a region where the physical and de Sitter Jacobi geometries are being matched.

For weak fields well inside the cosmological scale, Eq. \eqref{eq_FJM_desitter_compact_sec6} gives
\begin{equation}
F_{J,M}^{\rm dS}(r)
=
-\frac{2e^2-1}{2r(e^2-1)}
-\frac{\Lambda r(e^2+1)}{12(e^2-1)^2}
+
\mathcal{O}(\Lambda^2).
\label{eq_FJM_desitter_Lambda_expansion_sec6}
\end{equation}
We keep the de Sitter expansion in terms of the Killing energy $e$.  The substitution $e^2=(1-v^2)^{-1}$ in Eq. \eqref{eq_v_b_def_sec2} defines an asymptotic speed only after an asymptotically flat region is available, so we do not use it as a velocity parametrization inside the de Sitter static patch.  Applying the endpoint subtraction directly to the energy form above gives
\begin{align}
&\mathcal{P}^{\rm ren}_{J,{\rm Kot}}(r,r_\star)
=
r_g
\Bigg[
\frac{2e^2-1}{2(e^2-1)}
\left(
\frac{1}{r}-\frac{1}{r_\star}
\right)
\nn &+
\frac{\Lambda(e^2+1)}{12(e^2-1)^2}
(r-r_\star)
\Bigg]
+
\mathcal{O}(r_g\Lambda^2,r_g^2).
\label{eq_PJ_kottler_energy_expansion_sec6}
\end{align}
The first term is the Schwarzschild-type massive-particle contribution.  The second term is the leading mass-cosmological contribution generated by using the de Sitter Jacobi reference.

The primitive contribution to the finite-distance angle is evaluated along the reference trajectory at first order in $r_g$:
\begin{align}
&\mathcal{I}_{\rm Kot}^{(1)}
=
\int_{\phi_S^{(0)}}^{\phi_R^{(0)}}
\mathcal{P}^{\rm ren}_{J,{\rm Kot}}(r^{(0)}(\phi),r_\star)\,d\phi
\nn &=
r_g
\sum_{i=R,S}
\int_{r_0^{(0)}}^{r_i}
\left[
F_{J,M}^{\rm dS}(r_\star)-F_{J,M}^{\rm dS}(r)
\right]
\mathcal{R}_0(r,e,\ell)\,dr .
\label{eq_kottler_primitive_integral_sec6}
\end{align}
Here $r^{(0)}(\phi)$ denotes the de Sitter reference trajectory, whereas $r_0^{(0)}$ denotes its radial turning point.  This expression is the direct de Sitter-reference analogue of the asymptotically flat primitive integral.  Unlike the Minkowski case, it should not be separated from the reference and closure terms unless the same endpoint normalization is used consistently.

\subsection{Finite-distance bending and limiting regimes}
\label{subsec_kottler_finite_distance_limits}
The localized finite-distance bending is the first variation of the operational angle relative to the de Sitter reference at fixed $(e,\ell,r_S,r_R)$.  We write
\begin{equation}
\alpha_{J,{\rm Kot}}
=
\alpha_{J,{\rm dS}}
+
r_g\,\delta\alpha_{J,M}^{\rm dS}
+
\mathcal{O}(r_g^2).
\label{eq_kottler_alpha_expansion_sec6}
\end{equation}
The pure-reference term $\alpha_{J,{\rm dS}}$ is assigned to the no-localized-mass geometry.  Thus the lensing contribution of the central mass is $\delta\alpha_{J,M}^{\rm dS}$.

The endpoint part of this first variation follows from Eq. \eqref{eq_desitter_endpoint_angle_sec6}.  Defining
\begin{equation}
\sigma_i\equiv\sqrt{1-\chi_0(r_i)^2},
\qquad
i=R,S,
\label{eq_sigma_i_kottler_sec6}
\end{equation}
one obtains
\begin{equation}
\delta(\Psi_R-\Psi_S)_{M}^{\rm dS}
=
-\frac{1}{2}
\sum_{i=R,S}
\frac{\chi_0(r_i)}
{r_i\sigma_i}
\left[
\frac{1}{a(r_i)}
+
\frac{1}{W(r_i)}
\right].
\label{eq_kottler_endpoint_variation_sec6}
\end{equation}
The coordinate-span part may be written without expanding a singular lower limit by differentiating the exact regularized branch integrals.  If $r_0(r_g)$ is the physical turning point and $y=r_0(r_g)/r$, then
\begin{align}
&\delta\phi_{RS,M}^{\rm dS}
\nn &=
\left.
\frac{d}{dr_g}
\sum_{i=R,S}
\int_{r_0(r_g)/r_i}^{1}
\frac{r_0(r_g)}{y^2}
\mathcal{R}\!\left(\frac{r_0(r_g)}{y},e,\ell,r_g\right)dy
\right|_{r_g=0}.
\label{eq_kottler_phi_variation_regularized_sec6}
\end{align}
This representation keeps the turning point fixed at the upper endpoint $y=1$ for all $r_g$ and avoids spurious divergences from differentiating the square-root singularity naively.

The first-order localized bending is therefore
\begin{equation}
\delta\alpha_{J,M}^{\rm dS}
=
\delta(\Psi_R-\Psi_S)_{M}^{\rm dS}
+
\delta\phi_{RS,M}^{\rm dS}.
\label{eq_kottler_delta_alpha_operational_sec6}
\end{equation}
Equations \eqref{eq_kottler_endpoint_variation_sec6}-\eqref{eq_kottler_delta_alpha_operational_sec6} give the operational finite-distance expansion.  The Gauss-Bonnet form reorganizes the same quantity into the de Sitter reference contribution plus the discrepancy primitive of Eq. \eqref{eq_PJ_kottler_ren_sec6}.  The two forms agree when the same branch convention, endpoint normalization, and matching radius are used.

In the limit $\Lambda\to0$, the reference endpoint may be moved to infinity and $F_{J,M}^{\rm dS}(r_\star)\to0$.  Only at this asymptotically flat stage do we invoke Eq. \eqref{eq_v_b_def_sec2} and identify $e=(1-v^2)^{-1/2}$ with the asymptotic particle speed.  The primitive then becomes
\begin{equation}
\mathcal{P}^{\rm ren}_{J,{\rm Kot}}(r)
\longrightarrow
\frac{r_g}{2r}
\frac{2e^2-1}{e^2-1}
=
\frac{r_g}{2r}
\left(1+\frac{1}{v^2}\right).
\label{eq_kottler_schwarzschild_primitive_limit_sec6}
\end{equation}
Substitution into the Minkowski-reference trajectory gives
\begin{align}
\alpha_{J,{\rm Kot}}
&\longrightarrow
\frac{r_g}{2b}
\left(1+\frac{1}{v^2}\right)
\left[
\sqrt{1-\frac{b^2}{r_R^2}}
+
\sqrt{1-\frac{b^2}{r_S^2}}
\right]
\nn &+
\mathcal{O}\!\left(\frac{r_g^2}{b^2}\right).
\label{eq_kottler_schwarzschild_angle_limit_sec6}
\end{align}
With $r_g=2M$, this is precisely the finite-distance Schwarzschild massive-particle limit.

The ultra-relativistic limit inside the de Sitter reference is defined directly by $e\to\infty$.  From Eq. \eqref{eq_FJM_desitter_Lambda_expansion_sec6},
\begin{align}
\lim_{e\to\infty}F_{J,M}^{\rm dS}(r)
&=
-\frac{1}{r}
+
\mathcal{O}(\Lambda^2),
\nn
\lim_{e\to\infty}
\mathcal{P}^{\rm ren}_{J,{\rm Kot}}(r,r_\star)
&=
r_g\left(\frac{1}{r}-\frac{1}{r_\star}\right)
+
\mathcal{O}(r_g\Lambda^2).
\label{eq_kottler_null_limit_primitive_sec6}
\end{align}
After the asymptotically flat specialization, the same energy limit corresponds to $v\to1$.  Thus the leading optical mass contribution is recovered, while the pure de Sitter part remains assigned to the reference identity.

Finally, when $r_g=0$, the discrepancy density vanishes identically:
\begin{equation}
\Delta\mathcal{D}_J=0,
\qquad
\mathcal{P}^{\rm ren}_{J,{\rm Kot}}=0.
\label{eq_kottler_pure_desitter_zero_sec6}
\end{equation}
The formalism therefore gives no localized bending in pure de Sitter space.  This is the desired reference behavior: the cosmological background is not removed from the geometry, but it is used as the no-localized-mass standard relative to which the Kottler mass contribution is defined.

\section{Janis-Newman-Winicour naked-singularity application}
\label{sec_jnw_application}
We now use the Janis-Newman-Winicour spacetime as an application in which the reference-renormalized Jacobi primitive remains available even when circular-orbit normalization is not.  The geometry is asymptotically flat, so the reference is the ordinary Minkowski Jacobi metric at fixed asymptotic speed.  In the photon-sphere-free sector considered below, the circular timelike Jacobi orbit required by the orbit-normalized prescription can be absent for unbound scattering.  Our purpose is to show explicitly that primitive normalization does not require an auxiliary circular orbit.

\subsection{Geometry and obstruction to circular-orbit normalization}
\label{subsec_jnw_obstruction}
We use the convenient radial-coordinate form of the Janis-Newman-Winicour solution given by Virbhadra \cite{Virbhadra:1997ie}, who also showed its equivalence to Wyman's static massless-scalar solution.  In our notation,
\begin{align}
ds^2
&=
-f(r)^\gamma dt^2
+
f(r)^{-\gamma}dr^2
+
r^2 f(r)^{1-\gamma}d\Omega^2,
\nn
f(r)&=1-\frac{r_g}{\gamma r},
\label{eq_jnw_metric_sec7}
\end{align}
with $0<\gamma\le1$. The solution was originally obtained by Janis, Newman, and Winicour \cite{Janis:1968zz}.  For $0<\gamma<1$, the surface $r=r_g/\gamma$ is a curvature singularity without an event horizon.  Jhingan, Joshi, and Virbhadra showed that the JNW solution has a strong globally naked singularity \cite{Virbhadra:1995iy}.  Here \textit{strong} refers to the curvature singularity itself.  It is distinct from the gravitational-lensing classification of Virbhadra and Ellis \cite{Virbhadra:2002ju}, where weakly naked singularities possess a photon sphere and strongly naked singularities do not.  For JNW, $1/2<\gamma<1$ belongs to the weakly naked class, $0<\gamma<1/2$ belongs to the strongly naked class, and $\gamma=1/2$ is the limiting marginal case.  Virbhadra, Narasimha, and Chitre proposed scalar-field gravitational lensing as a possible observational means of distinguishing a naked-singularity lens from the Schwarzschild case \cite{Virbhadra:1998dy}.  Virbhadra and Keeton later found that strongly naked JNW lenses can produce images with negative time delays \cite{Virbhadra:2007kw}.  More recent lensing and imaging studies have further examined the optical behavior of JNW geometries \cite{Gyulchev:2019tvk,Gyulchev:2020cvo,Chen:2023uuy}.  The Schwarzschild limit is obtained for $\gamma=1$ and $r_g=2M$.  In the notation of Sec. \ref{sec_timelike_jacobi_observable},
\begin{align}
A(r)&=f(r)^\gamma,
\nn
B(r)&=f(r)^{-\gamma},
\nn
C(r)&=r^2f(r)^{1-\gamma},
\label{eq_jnw_functions_domain_sec7}
\end{align}
with $r>\frac{r_g}{\gamma}$. and $e^2>f(r)^\gamma$. The last two inequalities define the static and Riemannian Jacobi domain.

The interval $0<\gamma\le2/5$ used below lies strictly inside the strongly naked lensing class.  The upper bound $2/5$ comes from the timelike circular-orbit obstruction derived here rather than from the null photon-sphere threshold at $\gamma=1/2$.

A circular timelike Jacobi orbit at fixed energy must satisfy the general condition
\begin{equation}
\frac{d}{dr}
\left[
\frac{C(r)\left(e^2-A(r)\right)}{A(r)}
\right]_{r=r_{\rm co}}
=0 .
\label{eq_jnw_circular_condition_sec7}
\end{equation}
For the JNW functions, introduce $y=f(r)$, so that $0<y<1$ in the exterior singular domain.  Solving Eq. \eqref{eq_jnw_circular_condition_sec7} for the circular-orbit energy gives
\begin{equation}
e_{\rm co}^2(y)
=
y^\gamma
\frac{(1+\gamma)y+1-\gamma}
{(1+2\gamma)y+1-2\gamma}.
\label{eq_jnw_circular_energy_sec7}
\end{equation}
For $0<\gamma\le2/5$, the denominator is positive on $0<y<1$.  Differentiation gives
\begin{align}
\frac{d e_{\rm co}^2}{dy}
&=
\frac{\gamma y^{\gamma-1}Q_\gamma(y)}
{\left[(1+2\gamma)y+1-2\gamma\right]^2},
\nn
Q_\gamma(y)
&=
(1+3\gamma+2\gamma^2)y^2
-4\gamma^2y
+
(1-3\gamma+2\gamma^2).
\label{eq_jnw_circular_energy_derivative_sec7}
\end{align}
In this parameter range $Q_\gamma(y)>0$ for $0<y<1$, so $e_{\rm co}^2(y)$ is monotone increasing and tends to unity only as $y\to1$.  Hence
\begin{align}
&0<e_{\rm co}^2(y)<1,
\nn
&0<\gamma\le\frac{2}{5},
\nn
&0<y<1 .
\label{eq_jnw_no_unbound_circular_sec7}
\end{align}
Unbound scattering in the asymptotically flat normalization has $e\ge1$.  Therefore, throughout the photon-sphere-free JNW subinterval $0<\gamma\le2/5$ considered here, there is no circular timelike Jacobi orbit at the same energy as the unbound scattering particle.  The orbit-normalized primitive cannot be imposed in this sector.

\begin{figure}
    \centering
    \includegraphics[width=\columnwidth]{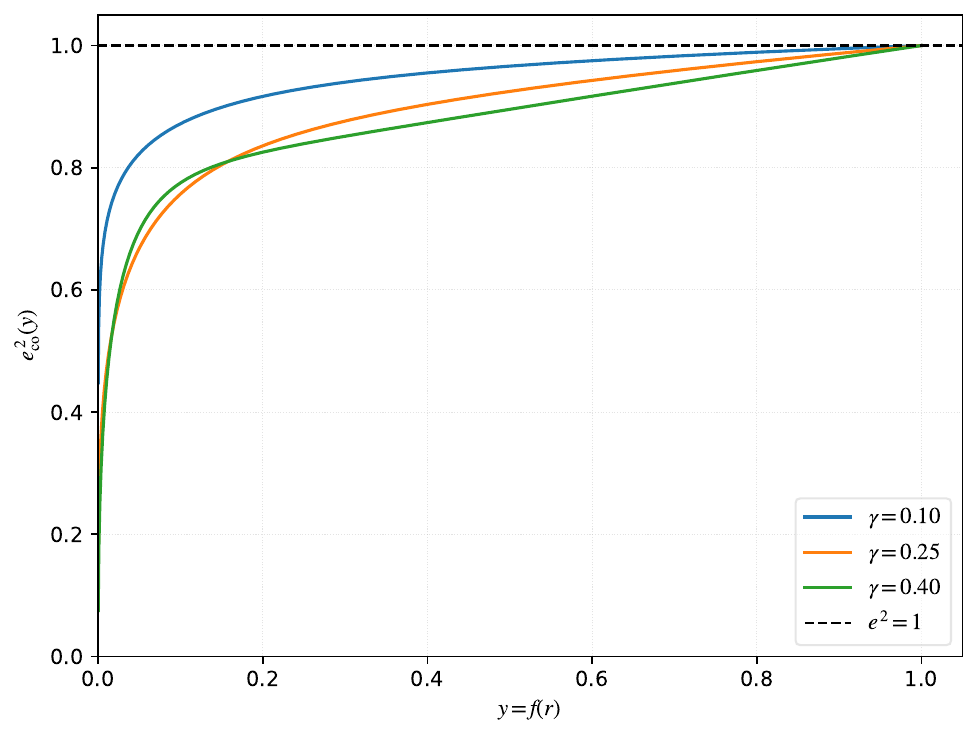}
    \caption{Circular-orbit obstruction in the photon-sphere-free JNW subinterval $0<\gamma\le2/5$.  The circular-orbit energy $e_{\rm co}^2(y)$ from Eq. \eqref{eq_jnw_circular_energy_sec7} is plotted as a function of $y=f(r)$ for representative values $0<\gamma\le 2/5$.  The dashed line marks the unbound-scattering threshold $e^2=1$.  For all displayed cases in this subinterval, $e_{\rm co}^2(y)$ remains below the threshold throughout $0<y<1$, confirming that no circular timelike Jacobi orbit exists at the same energy as an unbound scattering particle. The reference-renormalized primitive of Eq. \eqref{eq_PJ_jnw_exact_sec7} is therefore not replacing an available orbit normalization as it supplies a normalization where the circular-orbit prescription is unavailable.}
    \label{fig_jnw_circular_obstruction}
\end{figure}
Figure \ref{fig_jnw_circular_obstruction} visualizes the obstruction expressed analytically in Eqs. \eqref{eq_jnw_circular_energy_sec7}-\eqref{eq_jnw_no_unbound_circular_sec7}. In the photon-sphere-free JNW subinterval $0<\gamma\le2/5$ singled out by the timelike circular-orbit analysis, the circular-orbit energy remains below the unbound threshold $e^2=1$ for the entire exterior range $0<y<1$. Thus an unbound scattering particle cannot use a circular timelike Jacobi orbit as a primitive anchor in this sector.  This makes the JNW example a clean test of the reference-renormalized prescription: the endpoint-normalized primitive in Eq. \eqref{eq_PJ_jnw_exact_sec7} is well defined precisely in a regime where the circular-orbit normalization is not available.

\subsection{Reference-renormalized primitive and bending angle}
\label{subsec_jnw_primitive_bending}
The appropriate comparison geometry is the Minkowski Jacobi metric at the same asymptotic speed $v$, or equivalently at the same $e=(1-v^2)^{-1/2}$.  Since the spacetime is asymptotically flat, the matching endpoint is $r_\star=\infty$.  The curvature-density generator obtained from Eq. \eqref{eq_FJ_functional_sec4} is
\begin{align}
F^{\rm JNW}_J(r)
&=
\frac{r\sqrt{f(r)}}{2}
\Bigg[
\frac{2}{r}
+
\frac{(1-2\gamma)f'(r)}{f(r)}
\nn &-
\frac{\gamma f(r)^{\gamma-1}f'(r)}
{e^2-f(r)^\gamma}
\Bigg].
\label{eq_FJ_jnw_exact_sec7}
\end{align}
The prefactor $r\sqrt{f(r)}$ is the factor $\sqrt{C/B}$ specialized to the JNW metric.  Since $F^{\rm JNW}_J(\infty)=1$, the reference-renormalized primitive is
\begin{equation}
\mathcal{P}^{\rm ren}_{J,{\rm JNW}}(r)
=
1-F^{\rm JNW}_J(r).
\label{eq_PJ_jnw_exact_sec7}
\end{equation}
This equation is well defined independently of the existence of a circular orbit.

Expanding Eq. \eqref{eq_PJ_jnw_exact_sec7} for $r_g/r\ll1$ gives
\begin{align}
&\mathcal{P}^{\rm ren}_{J,{\rm JNW}}(r)
=
\frac{r_g}{2r}
\left(1+\frac{1}{v^2}\right)
+
\frac{r_g^2}{8\gamma^2r^2v^4}
\Big[
4\gamma^2v^2
-4\gamma^2
\nn & +2\gamma v^4
+2\gamma v^2
-v^4
\Big]
+
\mathcal{O}\!\left(\frac{r_g^3}{r^3}\right).
\label{eq_PJ_jnw_expansion_sec7}
\end{align}
The leading term depends only on the ADM mass scale $r_g/2$.  The naked-singularity parameter $\gamma$ first appears in the next term of the primitive expansion.

To complete the finite-distance angle through second order, we need the particle orbit and the angular endpoints only through first order.  We introduce the dimensionless quantities
\begin{equation}
\epsilon\equiv\frac{r_g}{b},
\qquad
z(\phi)\equiv\frac{b}{r(\phi)} .
\label{eq_jnw_epsilon_z_sec7}
\end{equation}
Substituting the JNW metric functions into Eq. \eqref{eq_orbit_equation_sec2} gives the exact first integral
\begin{equation}
\left(\frac{dz}{d\phi}\right)^2
=
f^{2-2\gamma}
\frac{e^2-f^\gamma}{e^2-1}
-z^2 f,
\qquad
f=1-\frac{\epsilon z}{\gamma}.
\label{eq_jnw_z_orbit_exact_sec7}
\end{equation}
Using $e^2=(1-v^2)^{-1}$ and expanding through first order in $\epsilon$, we obtain
\begin{equation}
\left(\frac{dz}{d\phi}\right)^2
=
1-z^2
+
\epsilon z
\left(
1+\frac{1}{v^2}-\frac{2}{\gamma}+\frac{z^2}{\gamma}
\right)
+
\mathcal{O}(\epsilon^2).
\label{eq_jnw_z_orbit_expansion_sec7}
\end{equation}
Differentiating Eq. \eqref{eq_jnw_z_orbit_expansion_sec7} away from the turning point and extending the result there by continuity gives the perturbative orbit equation.  We impose $z'(0)=0$ and use the turning-point condition at $\phi=0$ to fix the remaining homogeneous term.  The orbit through first order is
\begin{align}
z(\phi)
&=
\cos\phi
+
\epsilon
\left[
\mathcal{A}(v)
-
\frac{\cos^2\phi}{2\gamma}
\right]
+
\mathcal{O}(\epsilon^2),
\nn
\mathcal{A}(v)
&\equiv
\frac{1}{2}\left(1+\frac{1}{v^2}\right).
\label{eq_jnw_orbit_first_order_sec7}
\end{align}
For the finite endpoints, we define
\begin{equation}
z_i\equiv\frac{b}{r_i},
\qquad
s_i\equiv\sqrt{1-z_i^2},
\qquad
\theta_i\equiv\arccos z_i,
\label{eq_jnw_endpoint_symbols_sec7}
\end{equation}
and
\begin{equation}
\Phi_0\equiv\theta_R+\theta_S
=
\pi-\arcsin z_R-\arcsin z_S .
\label{eq_jnw_Phi0_sec7}
\end{equation}
Solving $z(\phi_i)=z_i$ with Eq. \eqref{eq_jnw_orbit_first_order_sec7} gives the physical angular endpoints through first order,
\begin{align}
\phi_R
&=
\theta_R
+
\epsilon\frac{\mathcal{A}(v)-z_R^2/(2\gamma)}{s_R}
+
\mathcal{O}(\epsilon^2),
\nn
\phi_S
&=
-\theta_S
-
\epsilon\frac{\mathcal{A}(v)-z_S^2/(2\gamma)}{s_S}
+
\mathcal{O}(\epsilon^2).
\label{eq_jnw_endpoint_angles_first_order_sec7}
\end{align}
Thus a second-order trajectory is not needed for the bending angle through $\mathcal{O}(\epsilon^2)$.  The first-order primitive multiplies the first-order orbit and endpoint shifts, while the second-order primitive is evaluated on the zeroth-order orbit.

We retain the second-order primitive coefficient already obtained in Eq. \eqref{eq_PJ_jnw_expansion_sec7} and write
\begin{equation}
\Xi_\gamma(v)
\equiv
\frac{
4\gamma^2v^2
-4\gamma^2
+2\gamma v^4
+2\gamma v^2
-v^4
}
{8\gamma^2v^4}.
\label{eq_Xi_gamma_sec7}
\end{equation}
Using Eq. \eqref{eq_jnw_orbit_first_order_sec7}, the primitive evaluated on the physical orbit becomes
\begin{align}
&\mathcal{P}^{\rm ren}_{J,{\rm JNW}}(r(\phi))
=
\epsilon\mathcal{A}(v)\cos\phi
\nn &+
\epsilon^2
\left\{
\mathcal{A}(v)
\left[
\mathcal{A}(v)-\frac{\cos^2\phi}{2\gamma}
\right]
+
\Xi_\gamma(v)\cos^2\phi
\right\}
\nn &+
\mathcal{O}(\epsilon^3).
\label{eq_jnw_primitive_on_orbit_second_order_sec7}
\end{align}
The angular integral that appears repeatedly is
\begin{equation}
\mathcal{J}_2
\equiv
\int_{-\theta_S}^{\theta_R}\cos^2\phi\,d\phi
=
\frac{1}{2}
\left(
\Phi_0+z_Rs_R+z_Ss_S
\right).
\label{eq_jnw_J2_sec7}
\end{equation}
The three second-order pieces then take the form
\begin{align}
\mathcal{I}^{(2)}_{\rm prim}
&=
\epsilon^2\Xi_\gamma(v)\mathcal{J}_2,
\nn
\mathcal{I}^{(2)}_{\rm orb}
&=
\epsilon^2\mathcal{A}(v)
\left[
\mathcal{A}(v)\Phi_0
-
\frac{\mathcal{J}_2}{2\gamma}
\right],
\label{eq_jnw_second_order_primitive_orbit_sec7}
\end{align}
and
\begin{equation}
\mathcal{I}^{(2)}_{\rm end}
=
\epsilon^2\mathcal{A}(v)
\sum_{i=R,S}
\frac{z_i}{s_i}
\left[
\mathcal{A}(v)-\frac{z_i^2}{2\gamma}
\right].
\label{eq_jnw_second_order_endpoint_sec7}
\end{equation}
Here $\mathcal{I}^{(2)}_{\rm prim}$ comes from the second-order primitive on the reference orbit, $\mathcal{I}^{(2)}_{\rm orb}$ comes from the first-order deformation of the orbit inside the first-order primitive, and $\mathcal{I}^{(2)}_{\rm end}$ comes from the first-order shifts of the angular endpoints.

Adding these pieces to the leading term gives the complete finite-distance bending angle through second order,
\begin{align}
&\alpha_{J,{\rm JNW}}
=
\frac{r_g}{b}\mathcal{A}(v)(s_R+s_S)
\nn
&+
\frac{r_g^2}{b^2}
\Bigg\{
\mathcal{A}(v)^2\Phi_0
+
\frac{1}{2}
\left[
\Xi_\gamma(v)-\frac{\mathcal{A}(v)}{2\gamma}
\right]
\nn & \times \left(
\Phi_0+z_Rs_R+z_Ss_S
\right)
\nn
&
+
\mathcal{A}(v)
\sum_{i=R,S}
\frac{z_i}{s_i}
\left[
\mathcal{A}(v)-\frac{z_i^2}{2\gamma}
\right]
\Bigg\}
+
\mathcal{O}\!\left(\frac{r_g^3}{b^3}\right).
\label{eq_jnw_full_second_order_sec7}
\end{align}
The parameter $\gamma$ therefore enters the finite-distance observable itself at second order.  The result also shows explicitly why the second-order primitive alone was insufficient.  The orbit and endpoint terms occur at the same order and must be retained together.

\subsection{Checks and interpretation}
\label{subsec_jnw_checks_interpretation}
For $\gamma=1$ and $r_g=2M$, Eq. \eqref{eq_jnw_full_second_order_sec7} reduces to the finite-distance Schwarzschild result through second order obtained in the massive-particle Jacobi analysis of Li and Jia \cite{Li:2019qyb},
\begin{align}
&\alpha_{J,{\rm JNW}}\big|_{\gamma=1}
=
\frac{M(1+v^2)}{bv^2}(s_R+s_S)
\nn &+
\frac{3M^2(4+v^2)}{4b^2v^2}\Phi_0
\nn
&+
\frac{M^2}{4b^2v^4}
\sum_{i=R,S}
\frac{z_i}{s_i}
\left[
3v^2(4+v^2)
+
(4-8v^2-3v^4)z_i^2
\right]
\nn &+
\mathcal{O}\!\left(\frac{M^3}{b^3}\right).
\label{eq_jnw_schwarzschild_limit_sec7}
\end{align}
This check includes the finite endpoint terms and therefore tests the orbit and endpoint corrections that enter the second-order result.

Sending both endpoints to infinity gives $z_R,z_S\to0$, $s_R,s_S\to1$, and $\Phi_0\to\pi$.  Equation \eqref{eq_jnw_full_second_order_sec7} then yields
\begin{align}
\alpha_{J,{\rm JNW}}^\infty
&=
\frac{r_g}{b}
\left(1+\frac{1}{v^2}\right)
+
\frac{\pi r_g^2}{16\gamma^2b^2v^2}
\nn & \times \left[
12\gamma^2+(4\gamma^2-1)v^2
\right]
+
\mathcal{O}\!\left(\frac{r_g^3}{b^3}\right).
\label{eq_jnw_infinite_distance_sec7}
\end{align}
For $\gamma=1$ and $r_g=2M$, the second term becomes $3\pi M^2(1+4/v^2)/(4b^2)$, in agreement with the Schwarzschild limit of Ref.~\cite{Li:2019qyb}.

The ultra-relativistic limit of the complete finite-distance result is
\begin{align}
&\alpha_{{\rm opt},{\rm JNW}}
=
\frac{r_g}{b}(s_R+s_S)
\nn
&+
\frac{r_g^2}{b^2}
\Bigg\{
\left(1-\frac{1}{16\gamma^2}\right)\Phi_0
\nn &+
\sum_{i=R,S}
\left[
\frac{z_i}{s_i}
\left(1-\frac{z_i^2}{2\gamma}\right)
-
\frac{z_i s_i}{16\gamma^2}
\right]
\Bigg\}
+
\mathcal{O}\!\left(\frac{r_g^3}{b^3}\right).
\label{eq_jnw_null_limit_sec7}
\end{align}
For infinite source and receiver distances, this becomes
\begin{equation}
\alpha_{{\rm opt},{\rm JNW}}^\infty
=
\frac{2r_g}{b}
+
\frac{\pi r_g^2}{b^2}
\left(1-\frac{1}{16\gamma^2}\right)
+
\mathcal{O}\!\left(\frac{r_g^3}{b^3}\right).
\label{eq_jnw_null_infinite_second_order_sec7}
\end{equation}
Setting $\gamma=1$ and $r_g=2M$ gives $4M/b+15\pi M^2/(4b^2)$, as required by the Schwarzschild optical limit.

The essential point is that Eqs. \eqref{eq_jnw_no_unbound_circular_sec7} and \eqref{eq_PJ_jnw_exact_sec7} address different questions.  Equation \eqref{eq_jnw_no_unbound_circular_sec7} states that, for $0<\gamma\le2/5$, the circular timelike Jacobi orbit required by the orbit-normalized prescription is unavailable at unbound scattering energy.  Equation \eqref{eq_PJ_jnw_exact_sec7} nevertheless defines a unique reference-renormalized primitive by matching to the Minkowski Jacobi geometry at infinity.  The finite-distance observable is therefore still computable, and the absence of a circular orbit is exposed as a limitation of one primitive gauge choice rather than a limitation of the Jacobi Gauss-Bonnet formulation itself.

\section{Discussion}
\label{sec_discussion}
The construction developed in this work separates two notions that are often combined in Jacobi Gauss-Bonnet calculations.  The first is the finite-distance scattering observable, defined by local source and receiver directions together with the coordinate angular span of the trajectory.  The second is the choice of additive representative for a radial curvature primitive.  Only the former is physical.  The latter is a normalization convention introduced to make the curvature-area term computable in a reduced one-dimensional form.  Circular-orbit normalization is one such convention, but it is not part of the definition of the bending angle.

Such a distinction is the main conceptual advantage of the reference-renormalized formulation.  In the exact Gauss-Bonnet identity, the curvature-area term depends on primitive differences.  A constant shift of the primitive therefore cancels before any observable is formed.  A circular orbit fixes this constant by requiring the primitive to vanish at a constant-radius Jacobi geodesic.  Reference renormalization fixes the same freedom differently: it subtracts a reference Jacobi geometry at the same conserved energy and imposes an endpoint normalization on the resulting curvature-density discrepancy.  When both constructions are admissible, they differ only by a constant and give the same exact curvature contribution.  When the circular orbit is unavailable, the reference prescription still gives a definite primitive provided the discrepancy is integrable toward the matching end.

The fixed-energy character of the subtraction is essential.  For photons in a static spacetime, the optical metric is determined by the spacetime geometry up to the conformal freedom irrelevant to null trajectories.  For massive particles, the Jacobi metric depends explicitly on the conserved energy.  The physical and reference Jacobi geometries must therefore be compared at the same value of the particle energy.  Otherwise, the subtraction would mix two different Maupertuis problems and would no longer isolate a gravitational or geometrical discrepancy at fixed scattering class. Because $e=-u_\mu\xi^\mu$, equality of the energy labels also presupposes a fixed relative normalization of the physical and reference static Killing generators.  In the Kottler application we implement this by obtaining the de Sitter member through $r_g\to0$ while retaining the same time coordinate and the same generator $\partial_t$. This is the main structural difference between the present massive-particle construction and its null optical analogue.

The reference geometry also has a second role in finite-distance problems.  It specifies the no-lens comparison geometry.  In asymptotically flat scattering this role is usually hidden, because the Minkowski reference has vanishing curvature density and the outer boundary contribution cancels the coordinate angular span in the standard way.  In conical or cosmological backgrounds, the reference cannot be treated as a disposable algebraic device.  The Schwarzschild-like bumblebee example shows that the massless geometry is flat but conical, so the reference closure contributes a finite term.  The Kottler example shows that the de Sitter member is the natural no-localized-mass geometry inside the static patch, so pure de Sitter terms belong to the reference identity rather than to the localized bending.  In both cases, the reference fixes how the finite-distance observable is decomposed.

This reference dependence should not be interpreted as an ambiguity in the exact observable.  The finite-distance angle defined by local directions and the trajectory span is fixed once the source, receiver, particle constants, and spacetime are fixed.  What changes with the reference is the decomposition of the Gauss-Bonnet identity into reference curvature, discrepancy curvature, and closure pieces.  At the exact level these pieces recombine into the same observable when treated consistently.  At finite perturbative order, a residual dependence on the matching convention can appear only through terms beyond the retained accuracy.  This is why the matching radius in a finite static patch should be chosen in a regime where the physical and reference Jacobi geometries are close in the adopted expansion.

The bumblebee application illustrates the compatibility with the circular-orbit approach in a setting where the reference is already non-Euclidean.  The localized primitive is a rescaled Schwarzschild-type primitive, but the full finite-distance angle contains a conical contribution that cannot be generated by the localized mass primitive alone.  The final expression separates the conical term from the finite-distance mass correction, and it reduces to the standard Schwarzschild massive-particle result when the Lorentz-violating deformation is removed.  The agreement with the circular-orbit Jacobi prescription follows because the two primitive normalizations are related by a constant shift.

The Kottler application illustrates a different aspect of the same method.  The cosmological term is not treated as a perturbation around Minkowski space, but as part of the reference geometry.  The localized mass parameter is then expanded relative to the de Sitter Jacobi metric at fixed conserved energy.  This organization is well adapted to finite-distance scattering inside the static patch, where the notions of asymptotic Euclidean straight line and asymptotic observer are not available in the usual sense. The fixed-energy comparison uses the common static Killing generator inherited from the Kottler-de Sitter family.  We introduce the asymptotic speed $v$ only after taking $\Lambda\to0$, when the asymptotically flat interpretation is available. The resulting mass-discrepancy primitive contains the Schwarzschild contribution in the limit of vanishing cosmological constant and organizes the leading mass-cosmological corrections in a way tied to the chosen de Sitter reference.

The JNW example shows that the reference-renormalized method is more than a reformulation of the circular-orbit prescription. In the photon-sphere-free JNW subinterval $0<\gamma\le2/5$ considered above, the circular timelike Jacobi orbit associated with unbound scattering energy is absent. Thus the orbit-normalized primitive cannot be imposed.  Nevertheless, the spacetime is asymptotically flat, and the discrepancy relative to the Minkowski Jacobi reference admits a well-defined endpoint-normalized primitive. The leading bending angle is set by the asymptotic mass scale, while the scalar parameter enters the complete finite-distance observable at second order in $r_g/b$. This example isolates the precise limitation of circular-orbit normalization: its failure is a failure of an auxiliary gauge choice, not of the finite-distance Jacobi Gauss-Bonnet construction.

Remarks about the limitations that remains, the present formulation is restricted to static, spherically symmetric spacetimes and to unbound single-pass trajectories.  These assumptions ensure that the equatorial Jacobi geometry is Riemannian on the full domain and that the projected orbit can be used as one side of a simply connected Gauss-Bonnet region.  Strong deflection, multiple winding, self-intersecting projected paths, and bound orbits require a different boundary decomposition.  In particular, bound timelike motion is naturally described by Jacobi geometry, but the relevant observable is an accumulated angular excess or precession rather than the source-to-receiver bending angle used here.

A further limitation is the static assumption.  For stationary axisymmetric spacetimes, the massive-particle Jacobi-Maupertuis structure is generally Randers-Finsler rather than Riemannian.  The radial total-derivative primitive used in the present work relies on spherical symmetry and on the existence of a diagonal two-dimensional Jacobi metric depending only on one radial variable.  Extending the reference-renormalized idea to rotating geometries would require either a carefully controlled osculating Riemannian construction or a direct Finslerian treatment of the boundary and geodesic-curvature terms.

The same caution applies to media and external forces.  Plasma lensing, dispersive optical geometries, and charged-particle motion in external electromagnetic fields may admit effective optical or Jacobi-type descriptions, but the parameter held fixed in the reference subtraction changes from case to case.  For massive geodesic motion, the relevant fixed parameter is the conserved energy per unit rest mass.  For dispersive media it may be frequency; for charged particles it may involve both energy and charge-to-mass ratio.  A reference-renormalized primitive can be meaningful in such systems only if the physical and reference variational problems are compared at the same operationally conserved quantities.

The broader lesson is that primitive normalization should be chosen to match the physical comparison problem rather than imposed through an auxiliary orbit whenever such an orbit happens to exist.  Circular orbits remain useful and, in many black-hole examples, provide an elegant shortcut.  The reference-renormalized prescription retains that shortcut in the overlap regime but removes it as a prerequisite.  This makes the Jacobi Gauss-Bonnet method applicable to finite-distance massive-particle scattering in geometries where the natural comparison background is conical, cosmological, or lacking an admissible circular timelike anchor.

\section{Conclusions}
\label{sec_conclusions}
We have developed a reference-renormalized Jacobi-primitive Gauss-Bonnet formalism for finite-distance weak deflection of massive particles in static, spherically symmetric spacetimes.  The construction begins with the fixed-energy Jacobi metric associated with timelike geodesics and uses the Gauss-Bonnet theorem to represent the finite-distance bending angle through curvature and boundary contributions on the equatorial Jacobi manifold.  The central technical observation is that the radial curvature density is a total derivative, so the curvature-area term may be reduced to a primitive evaluated along the projected trajectory.  Since this primitive is defined only up to an additive constant, its normalization must be specified.  Instead of imposing a vanishing condition at a circular timelike orbit, we fix the primitive by subtracting a reference Jacobi geometry at the same conserved particle energy and by imposing endpoint normalization on the curvature-density discrepancy.

We have shown that circular-orbit normalization is not required of the finite-distance Jacobi Gauss-Bonnet formulation.  It is an additive gauge choice for the curvature primitive.  When an admissible circular timelike orbit exists in the same regular Jacobi domain, the circular-orbit-normalized primitive and the reference-normalized primitive differ only by a constant, and this constant cancels from the exact curvature strip integral.  The two prescriptions therefore agree in their common domain of validity.  The advantage of the reference-renormalized prescription is that it remains meaningful when the circular orbit is absent, lies outside the Riemannian Jacobi region, or is not compatible with the finite-distance reference geometry used to define the no-lens comparison.

The applications demonstrate the utility of the construction without relying on a standalone Schwarzschild calculation.  In the Schwarzschild-like bumblebee spacetime, the method separates the conical reference contribution from the localized mass correction and reproduces the circular-orbit Jacobi result after translating conventions; the usual Schwarzschild massive-particle bending is recovered by taking the conical parameter to unity.  In the Kottler spacetime, the de Sitter Jacobi geometry provides the natural fixed-energy reference inside the static patch, and the localized mass contribution is organized through a de Sitter discrepancy primitive. The physical and reference energies use the same normalization of the static Killing generator, while the speed $v$ is reserved for the asymptotically flat limit. The Schwarzschild and optical limits are recovered in the appropriate regimes.  In the Janis-Newman-Winicour spacetime, the photon-sphere-free subinterval $0<\gamma\le2/5$ selected by our timelike circular-orbit analysis provides a regime in which unbound circular-orbit normalization is unavailable, while the reference-renormalized primitive remains well defined relative to the Minkowski Jacobi reference. The finite-distance angle can then be carried through second order, where the scalar parameter first enters the observable. This example makes explicit that the absence of a circular timelike anchor obstructs only one primitive normalization, not the finite-distance Jacobi Gauss-Bonnet method itself.

Several extensions are natural. Higher-order finite-distance expansions can be developed by combining the endpoint-angle expansion, turning-point shift, and reference-renormalized primitive beyond the perturbative orders retained in the applications. Stationary axisymmetric spacetimes require a separate treatment because massive-particle Jacobi-Maupertuis geometry becomes Randers-Finsler rather than Riemannian, and the boundary terms must be handled with greater care.  Other extensions include dispersive media, charged-particle motion, finite static patches with different reference geometries, and strong-deflection domains with multiple winding or nontrivial boundary decomposition.  The guiding principle in all such cases is the same: the primitive normalization should be tied to the physical reference problem and to the parameters held fixed in the variational principle, not to the accidental existence of an auxiliary circular orbit.

\acknowledgments
R. P. would like to acknowledge networking support of the COST Action CA21106 - COSMIC WISPers in the Dark Universe: Theory, astrophysics and experiments (CosmicWISPers), the COST Action CA22113 - Fundamental challenges in theoretical physics (THEORY-CHALLENGES), the COST Action CA21136 - Addressing observational tensions in cosmology with systematics and fundamental physics (CosmoVerse), the COST Action CA23130 - Bridging high and low energies in search of quantum gravity (BridgeQG), and the COST Action CA23115 - Relativistic Quantum Information (RQI) funded by COST (European Cooperation in Science and Technology). R. P. also thanks SCOAP3 (Switzerland) for their support.

\bibliography{ref}

\end{document}